\documentclass[preprint,superscriptaddress]{revtex4-1}
\usepackage{url}
\usepackage{graphicx}
\usepackage{amsmath}
\usepackage{color}
\usepackage{multirow}
\usepackage{longtable}
\usepackage{array}
\usepackage{soul}
\newcolumntype{L}[1]{>{\raggedright\let\newline\\\arraybackslash\hspace{0pt}}m{#1}}
\newcolumntype{C}[1]{>{\centering\let\newline\\\arraybackslash\hspace{0pt}}m{#1}}
\newcolumntype{R}[1]{>{\raggedleft\let\newline\\\arraybackslash\hspace{0pt}}m{#1}}
\begin{document}

\title{Characterization of Spatial Coherence of Synchrotron Radiation with Non-Redundant Arrays of Apertures}

\author{P. Skopintsev}
\affiliation{Deutsches Elektronen-Synchrotron DESY, Notkestrasse 85, D-22607 Hamburg, Germany}
\affiliation{National Research Center "Kurchatov Institute", Kurchatov square 1, 123182 Moscow, Russia}
\author{A. Singer}
\altaffiliation{present address: University of California, San Diego, CA 92093, USA}
\affiliation{Deutsches Elektronen-Synchrotron DESY, Notkestrasse 85, D-22607 Hamburg, Germany}
\author{J. Bach}
\affiliation{Universit\"at Hamburg, Institut f\"ur Angewandte Physik, D-20355 Hamburg, Germany}
\author{L. M\"uller}
\affiliation{Deutsches Elektronen-Synchrotron DESY, Notkestrasse 85, D-22607 Hamburg, Germany}
\author{B. Beyersdorf}
\affiliation{Universit\"at Hamburg, Institut f\"ur Angewandte Physik, D-20355 Hamburg, Germany}
\author{S. Schleitzer}
\affiliation{Deutsches Elektronen-Synchrotron DESY, Notkestrasse 85, D-22607 Hamburg, Germany}
\author{O. Gorobtsov}
\affiliation{Deutsches Elektronen-Synchrotron DESY, Notkestrasse 85, D-22607 Hamburg, Germany}
\affiliation{National Research Center "Kurchatov Institute", Kurchatov square 1, 123182 Moscow, Russia}
\author{A. Shabalin}
\affiliation{Deutsches Elektronen-Synchrotron DESY, Notkestrasse 85, D-22607 Hamburg, Germany}
\author{R. Kurta}
\affiliation{Deutsches Elektronen-Synchrotron DESY, Notkestrasse 85, D-22607 Hamburg, Germany}
\author{D. Dzhigaev}
\affiliation{Deutsches Elektronen-Synchrotron DESY, Notkestrasse 85, D-22607 Hamburg, Germany}
\affiliation{National Research Nuclear University, "MEPhI", 115409 Moscow, Russia}
\author{O.M. Yefanov}
\affiliation{Deutsches Elektronen-Synchrotron DESY, Notkestrasse 85, D-22607 Hamburg, Germany}
\author{L. Glaser}
\affiliation{Deutsches Elektronen-Synchrotron DESY, Notkestrasse 85, D-22607 Hamburg, Germany}
\author{A. Sakdinawat}
\affiliation{SLAC National Accelerator Laboratory, 2575 Sand Hill Road, Menlo Park, California 94025, USA}
\author{Y. Liu}
\affiliation{University of California, Berkeley, California 94720, USA}
\author{G. Gr\"ubel}
\affiliation{Deutsches Elektronen-Synchrotron DESY, Notkestrasse 85, D-22607 Hamburg, Germany}
\affiliation{The Hamburg Center for Ultrafast Imaging, D-22761 Hamburg, Germany}
\author{R. Fr\"omter}
\author{H.P. Oepen}
\affiliation{Universit\"at Hamburg, Institut f\"ur Angewandte Physik, D-20355 Hamburg, Germany}
\author{J. Viefhaus}
\affiliation{Deutsches Elektronen-Synchrotron DESY, Notkestrasse 85, D-22607 Hamburg, Germany}
\author{I.A.Vartanyants}
\affiliation{Deutsches Elektronen-Synchrotron DESY, Notkestrasse 85, D-22607 Hamburg, Germany}
\affiliation{National Research Nuclear University, "MEPhI", 115409 Moscow, Russia}

\date{\today}

%\keyword{Coherence, Interference, Synchrotron X-ray sources}

%\maketitle

\begin{abstract}
We present a method to characterize the spatial coherence of soft X-ray radiation from a single diffraction pattern.
The technique is based on scattering from non-redundant arrays (NRA) of slits and records the degree of spatial coherence at several relative separations from one to 15 microns, simultaneously.
Using NRAs we measured the transverse coherence of the X-ray beam at the XUV X-ray beamline P04 of the PETRA III synchrotron storage ring as a function of different beam parameters.
To verify the results obtained with the NRAs additional Young's double pinhole experiments were conducted and show good agreement.
\end{abstract}
\maketitle

\section{Introduction}
It is well established that the
present-day third generation synchrotrons are partially coherent sources \cite{Winick,Attwood,VartanyantsNJP2010}.
With the construction of these facilities, new research areas utilizing partial coherence of X-ray radiation have emerged.
The most prominent experimental techniques are coherent X-ray diffractive imaging (CXDI) \cite{Vartanyants2010,ChapmanNatPhot2010,MancusoNJP2010}, X-ray holography \cite{Eisebitt_Nature_2004,Stickler_APL_2010} and X-ray photon correlation spectroscopy (XPCS) \cite{GruebelJAC2004}.
In CXDI static real space images of the sample are obtained by phase retrieval techniques \cite{FienupApplOpt1982}, while correlation techniques are applied in XPCS to explore system dynamics \cite{G2007}.
The key feature of the methods utilizing the high degree of coherence is the interference of the field scattered by different parts of the sample under study.
Hence, spatial coherence across the sample is essential and understanding the coherence properties of the beams at new generation X-ray sources is of crucial importance for the scientific community.
Moreover, a detailed knowledge of the beam coherence can be used to improve the resolution obtained in the CXDI phase retrieval \cite{WhiteheadPRL2009}.

The Young's double pinhole experiment is the most direct technique to map out the spatial coherence. The method has been successfully applied for soft X-ray synchrotron radiation \cite{ChangOptCommun2000,NugentSynchr2001} and, recently, for free-electron lasers (FELs) \cite{Singer_FLASH_2008,Singer_LCLS_2011,Singer_FLASH_2012}.
However, a single double pinhole experiment yields the spatial coherence at only one relative distance, the distance between the pinholes.
To fully characterize the spatial coherence several measurements at different pinhole separations are required.
Other techniques have been implemented to measure the spatial coherence, such as X-ray grating interferometry \cite{Pfeiffer_grating_2005}, phase space tomography \cite{Raymer_tomogr_1994,Tran_tomogr_2007}, methods which utilize scattering on Brownian particles \cite{AlaimoPRL2009}, and measurements of the intensity correlations \cite{GluskinJSynchRad1999,YabashiPRL2001,Singer_HBT_2013}.
Similar to the double pinhole experiment, the aforementioned techniques require a number of measurements to fully determine the spatial coherence properties of the probing radiation.

Here we present a method to completely characterize the spatial coherence function of the soft X-ray radiation from a single diffraction pattern.
This method utilizes a non-redundant array (NRA), that is a mask with multiple slits where each slit separation appears only once.
Each pair of slits acts as a single Young's interferometer, and a measurement using an NRA can be considered as a series of double pinhole interferograms acquired simultaneously.
The signal from different slit pairs can be discriminated due to the uniqueness of each slit separation.
This is in contrast to coherence measurements using a uniformly redundant array (URA), where each slit separation appears more than once \cite{LinPRL2003}.
The analysis of URA measurements is quite sensitive to the wavefront of the incident radiation \cite{Vartanyants2010}, a limitation that is not present in NRA measurements.
NRAs were previously used to measure spatial coherence of He-Ne laser \cite{MejiaOptCommun2007,GonzalezJOSA2011}.
%We anticipate that this method is a viable and efficient technique to characterize spatial coherence at synchrotron sources, and will be especially interesting and useful at FEL sources, as it will allow to characterize single FEL pulses.
In this paper we demonstrate spatial coherence measurements performed at the XUV X-ray beamline P04 at PETRA III using NRA apertures.

\section{Theory}

In the theory of optical coherence, the statistical properties of the radiation, including spatial coherence, are described by correlation functions of the electric field.
The mutual coherence function (MCF) is defined as \cite{G1985}
% is a measure directly related to interference phenomena and describing correlation between two complex values
%    of the electric field $E_{1}, E_{2}$ at two points \textbf{\emph{$r_1$}}, \textbf{\emph{$r_2$}} in space and time $t$, separated by time interval $\tau$. By definition, it is expressed by \cite{G1985}
    \begin{equation}
    \Gamma(\mathbf r_1,\mathbf r_2;\tau) = \langle E^*(\mathbf r_1,t) E (\mathbf r_2 ,t + \tau) \rangle
    \end{equation}
where $E(\mathbf r_1,t)$ and $E(\mathbf r_2,t+\tau)$ are the field values at the positions $\mathbf r_1$ and $\mathbf r_2$ and times $t$ and $t+\tau$ and $\langle \cdots \rangle$ is the ensemble average.
The intensity $I(\mathbf r)$ and complex degree of coherence $\gamma_{12}(\tau)$ can be readily obtained from the MCF
    \begin{equation}
    	I(\mathbf r) = \Gamma(\mathbf r,\mathbf r; 0),\qquad
	\gamma_{12}(\tau) = \frac{ \Gamma(\mathbf r_1,\mathbf r_2;\tau)}{\sqrt{I(\mathbf r_1)\cdot I(\mathbf r_2)}}.
    \label{eq:Eq1}
    \end{equation}
If the time delay $\tau$ is much smaller than the coherence time $\tau_c$, $\gamma_{12}(\tau)$ can be well approximated by the complex coherence factor (CCF) $\gamma_{12} = \gamma_{12}(0)$ \cite{G1985} that does not depend on the time delay $\tau$.

In the frame of the Gaussian Schell-model, which in most cases provides sufficient description of synchrotron radiation \cite{VartanyantsNJP2010}, the intensity profile and the CCF are both Gaussian functions. The beam size in this model is characterized by its root mean square (rms) width $\sigma$ of the intensity and the transverse coherence length $l_c$ can be defined as the rms width of $|\gamma_{12}|$. To characterize the transverse coherence by one number the global degree of coherence can be introduced \cite{SaldinOptCommun2008,VartanyantsNJP2010}
\begin{equation}
\zeta = \frac{(l_c / \sigma)}{\sqrt{4 + (l_c/\sigma)^2}}.
\end{equation}

%To experimentally characterize the CCF, the Young's double pinhole experiment can be used, where interference between the fields transmitted through pinhole one and two is observed.
%The degree of coherence can be determined from the contrast of the interference fringes.
%The coodinate $q$ is momentum transfer due to field scattering on the pinhole and $\tau$ is the time delay, which is introduced through the difference between the propagation path lengths of the two beams.
%In general approach, multiple apertures of any desirable shape could be considered in such an experiment.

\begin{figure}
        \centering
        \includegraphics[width = 1.0\columnwidth]{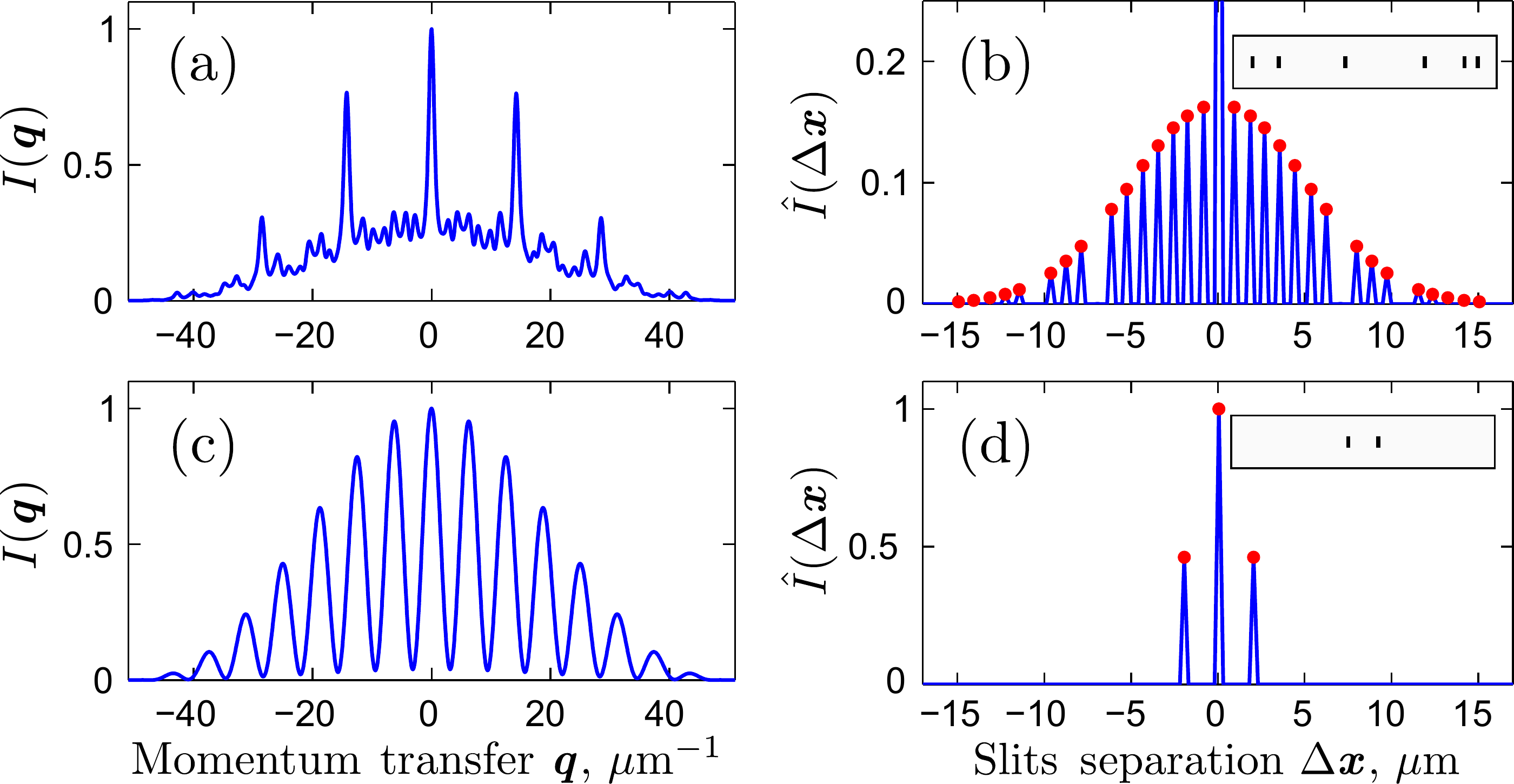}
    \caption{(a, c) Simulated diffraction patterns $I(q)$ from NRAs of six (a) and two (c) slits.
    (b, d) Fourier transform $\hat{I}(\Delta x)$ of corresponding diffraction patterns from six (b) and two (d) slits.
    Red points in (b, d) denote the central and satellite peaks maxima $C_0$ and $C_{ij}$, respectively. The insets in (b) and (d) show the slits relative positions used in simulations.
    The radiation was uniformly distributed in the slits and had a coherence length of $l_c=5$ $\mu$m.}
    \label{fig:Figure1}
    \end{figure}

To measure the CCF of the radiation an arrangement of apertures can be used as a scattering object.
The correlations between the field scattered at the apertures and propagated to the detection plane can be recorded as intensity modulations or interference fringes.
For a narrow bandwidth radiation, the far-field diffraction pattern $I(q)$ of $N$ identical apertures can be expressed in the following way  \cite{MejiaOptCommun2007,GonzalezJOSA2011}
    \begin{equation}
    %I(q) = I_S(q) \bigg( C_0 + 2\sum_{i \neq j}^N C_{ij} \cos \big( q \cdot d_{ij} + \alpha_{ij} \big) \bigg),
    I(\mathbf q) = I_S(\mathbf q) \cdot\bigg( C_0 +  \sum_{i \neq j}^N C_{ij}e^{i (\mathbf q\cdot \mathbf d_{ij}+\alpha_{ij})} \bigg),
    \label{eq:Eq2}
    \end{equation}  % big formula -> might be "bad box"
where coefficients $C_0$ and $C_{ij}$ are defined as
\begin{equation}
C_0=\sum_{i} I_i , \qquad C_{ij} = |\gamma_{ij}| \sqrt{I_i I_j} .
\label{eq:Eq5}
\end{equation}
Here $I_i$ is the intensity incident on the aperture $i$ \footnote{Here we assume that the incident intensity is constant across a single aperture}, $|\gamma_{ij}|$ is the modulus of the CCF at the aperture separation $\mathbf d_{ij}=-\mathbf d_{ji}$, $\alpha_{ij}=-\alpha_{ji}$ is the relative phase, $I_S(\mathbf q)$ is the intensity distribution on the detector due to beam diffraction on a single aperture, and $\mathbf q$ is the momentum transfer vector.
If the aperture has a form of a pinhole, then $I_S(\mathbf q)$ is the well-known Airy pattern. For a rectangular slit, $I_S(\mathbf q)$ has a shape of a squared sinc function.
An example of the intensity distribution $I(\mathbf q)$ for diffraction on six and two slits is shown in Fig. \ref{fig:Figure1}(a, c).
The interference pattern in equation \eqref{eq:Eq2} can be conveniently analyzed using the Fourier transform $\Hat{I}(\Delta \mathbf x)$ of $I(\mathbf q)$, which for $N$ apertures is given by \cite{BartelsScience2002,MejiaOptCommun2007,GonzalezJOSA2011}
    \begin{equation}
    \Hat{I}(\Delta \mathbf x) = \Hat{I_S}(\Delta \mathbf x) \otimes \bigg[ C_0 \delta (\Delta \mathbf x) + \sum_{i \neq j}^N C_{ij}\big\{ e^{i \alpha_{ij}} \delta ( \Delta \mathbf x - \mathbf d_{ij} ) \big\} \bigg],
   \label{eq:Eq3}
   \end{equation}
where $\delta (\Delta \mathbf x)$ is the Dirac delta function, $\Hat{I}_S(\Delta \mathbf x)$ is the Fourier transform of $I_S(\mathbf q)$, the symbol $\otimes$ denotes the convolution, and $\Delta \mathbf x$ denotes the relative distance.
It is clear from expression \eqref{eq:Eq3} that the peak maxima at each peak separation $\mathbf d_{ij}$ is given by coefficients $C_{ij}$.
If the intensities $I_i$ incident on the apertures are known, the CCF can be readily obtained from equation \eqref{eq:Eq5} as
   \begin{equation}
        |\gamma_{ij}| = C_{ij}/\sqrt{I_i I_j}, \\
   \label{eq:Eq4}
   \end{equation}
for all $d_{ij}$.
An NRA with $N$ slits shows $N(N-1)+1$ individual peaks in the Fourier transform $\Hat{I}(\Delta \mathbf x)$ of the measured intensity $I(\mathbf q)$, whereas each peak corresponds to interference between a unique combination of two slits.
According to equation \eqref{eq:Eq4} each peak in equation \eqref{eq:Eq3} corresponds to a unique value of $|\gamma_{ij}|$.
A typical distribution $\hat I(\Delta \mathbf x)$ with 31 and 3 peaks, corresponding to six and two slit apertures are shown in Fig. \ref{fig:Figure1}(b, d).

%As each $|\gamma_{ij}|$ corresponds to particular satellite peak, the modulus of CCF function can be determined on $N(N-1)/2$ distances.

It is clear that the outcome of the measurement strongly depends on the design of the NRA.
To optimize the NRA structure one possibility would be to find a uniform spatial distribution of the peaks.
It can be shown that such an ideal NRA with a uniform distribution of the peaks can only be achieved with four slits \cite{GolombProof}.
For more than four slits the peaks can be uniformly distributed, but some of them will be missing.
However, it is still possible to find an arrangement of N slits with as few missing peaks as possible.
In this work we designed NRAs using a so-called Golomb ruler
\footnote{A list of all today available optimal Golomb rulers can be found in Ref. \cite{Golomb} (see also \cite{wiki:Golomb}).}
that has integer marks with a distinct distance between every two marks.

%In mathematics, a so called Golomb ruler has integer marks with a distinct distance between every two marks
%and is identical to an NRA.
%Golomb Rulers find diverse applications in computer science and electrical engineering and the so called optimal Golomb rulers possess exactly the property described above

% The most direct way to obtain $I_i$ is to measure beam profile $I(r)$ at the apertures plane $r$. In this case, the intensity at the apertures is $I_i = I(r_i)$, where $r_i$ is the aperture $i$ coordinate. When lacking accurate beam cross-section intensity measurements, one could use a model Gaussian beam approximation
%
%\begin{equation}
% I_i = e^{-\frac{(r_i - r_0)^2}{2 \sigma^2}},
% \label{eq:Eq8}
%\end{equation}
%\\
%where $\sigma$ is a coefficient proportional to beam full width at half maximum ($FWHM = 2 \sqrt{2 ln(2)}\sigma$) and $r_0$ is the relative shift of the beam.

% PUT $|\gamma_{12}| = |\gamma_{\Delta r}|$ somewhere

\section{Experiment}

The coherence measurements were performed at the Variable Polarization XUV beamline P04 {\cite{JensP04_2013} of the PETRA III synchrotron radiation source at DESY in Hamburg during comissioning.
The beamline setup is schematically presented in Fig. \ref{fig:Figure2}.
A five meter APPLE-II type helical undulator with 72 magnetic periods was tuned to deliver 400 eV (wavelength $\lambda = 3.1$ nm) photons.
The beam further propagated to the end-station through several optical elements, including a beam defining slit (27 m downstream from the undulator), horizontal plane mirror (35 m), and vertical plane mirror together with a plane varied-line-spacing (VLS) grating (46 m).
The VLS grating focused the beam at the exit slit (71 m).
An elliptical mirror (78.5 m) focused the beam vertically to the sample position (81 m).
Horizontally the beam was collimated by a cylindrical mirror (79.1 m) and defined by a slit in front of the sample.
All mirrors were designed to accept $6\sigma$ rms of the beam size.
\begin{figure}
        \centering
        \includegraphics[width = 1.0\columnwidth]{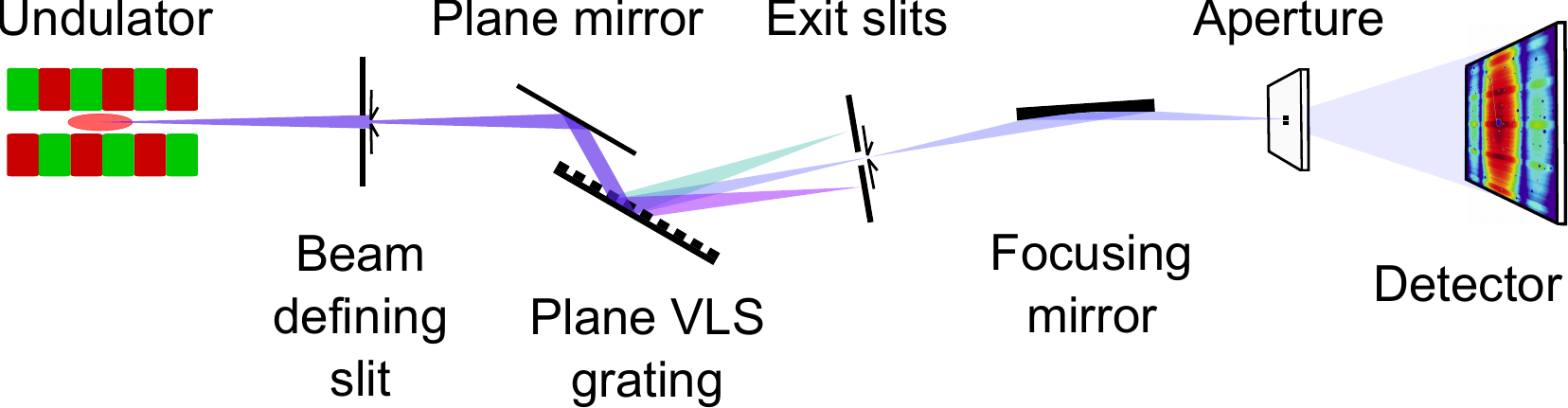}
    \caption{Experimental setup.
    Synchrotron radiation is generated by a 5 m APPLE-II type undulator and is transmitted through the beamline including a beam defining slit, monochromator comprised of a VLS plane grating and pair of plane mirrors, exit slit, focusing mirror, and collimating mirror.
    The beam is focused on the apertures by the focusing mirror and forms a diffraction pattern on the CCD detector positioned in the far-field. The horizontal plane mirror and the collimating mirror are not shown in the figure.}
    \label{fig:Figure2}
    \end{figure}
NRAs and double pinholes were manufactured in an opaque screen (see \cite{VartanyantsPRL2011} for a detailed description of the manufacturing process) and were used to map out the transverse coherence of the beam.
The distance between individual slits in the NRA aperture was varying between 1 $\mu$m and 15 $\mu$m.
The slits were positioned at -7.5 $\mu$m, -6.62 $\mu$m, -3.97 $\mu$m, 1.32 $\mu$m, 5.74 $\mu$m, and 7.5 $\mu$m and each slit was 0.8 x 0.25 $\mu$m$^2$ in size with the smaller size being along the NRA direction.
Double pinhole apertures had a separation between 2 $\mu$m and 15 $\mu$m, whereas the pinhole size varied between 0.34 $\mu$m and 0.5 $\mu$m.
The apertures were mounted on a piezo-positioner stage with a travel range of up to 20 mm.
A position uncertainty of $\pm 1$ $\mu$m  was observed during the experiment.
Finally, the interference patterns were recorded by "Spectral Instruments" 1100S charge-coupled device (CCD) with 4096$\times$4112 pixels, each 15$\times$15 $\mu$m$^2$ in size.
The detector was positioned 0.8 m downstream of the sample and a round beamstop 700 $\mu$m in size protected the detector from the direct beam.

The transverse coherence and the intensity profile in the vertical direction were measured for different beamline parameters, including the monochromator exit slit, beam defining slit downstream of the undulator, and the photon energy transmitted through the monochromator.
The monochromator exit slit width was varied between 40 $\mu$m and 230 $\mu$m, while the monochromator was tuned to an energy of 400 eV (maximum flux) and the beam defining slit was opened to 4.7 mm.
In this initial stage of beamline commissioning the photon energy resolution was the same for both exit slit openings, resulting in a bandwidth of about 0.5 eV \footnote{At this very early stage of commissioning the grating did not yet deliver the designed resolving power ($E/\Delta E > 10000$)}.
%The photon energy resolution was the same for both exit slits opening, having a bandwidth of about 0.5 eV at 400 eV.
This gave us an estimate of the temporal coherence length $l_t=c\cdot\tau_c$ that was $l_t=1.6$ $\mu$m.
%and corresponds to $c\cdot\tau_c/\lambda \approx 531$ wavelengths.
For the rest of the measurements the exit slit size was fixed to the value $D_\textrm{es}=230$ $\mu$m and the beam defining slit size $D_\textrm{ds}$ was chosen to be 4.7 mm, 1.7 mm, and 0.8 mm.
These slits correspond to a slit transmission of 1, 0.5, and 0.25, respectively.
For each beam defining slit width we measured the transverse coherence at photon energies of 396.5 eV, 400 eV and 403 eV, which were selected by the monochromator for a fixed undulator gap.
At 400 eV the maximum flux was observed, whereas at 396.5 eV and 403 eV the flux was reduced by a factor of 2.

%Hence, temporal coherence length $c\cdot\tau_c = \sqrt{\frac{2 ln(2)}{\pi}}\cdot\frac{1}{\Delta \nu}$ is approximately 1.6 $\mu$m \cite{G1985}.

Six slits NRA diffraction patterns were collected for all beamline parameters.
Additionally, double pinholes interferograms were measured for the case of
exit slit width of 40 $\mu$m and 230 $\mu$m.
A sufficient sampling of interference fringes was obtained in all cases with at least 11 pixels per fringe for a 15 $\mu$m apertures separation.

\section{Results and discussion}

Typical recorded and dark field corrected diffraction patterns of six slits NRA and double pinholes are shown in Figs. \ref{fig:Figure3}(a) and \ref{fig:Figure3}(c), respectively. Both interferograms were measured at 400 eV photon energy, 4.7 mm beam defining slits opening, and the monochromator exit slits width $D_\textrm{es} = 230$ $\mu$m.
Peaks corresponding to the apertures interference are clearly visible in the Fourier transforms of both diffraction patterns presented in Figs. \ref{fig:Figure3}(b) and \ref{fig:Figure3}(d).
It is interesting to note that in these Fourier transforms of diffraction patterns the contribution of the higher harmonic radiation at 800 eV photon energy transmitted through the monochromator reveals itself in additional peaks at larger distances visible in Fig. \ref{fig:Figure3}(d).
%The outer peaks at larger distances visible in Fig. \ref{fig:Figure3}(d) are due to the higher harmonic radiation at 800 eV photon energy transmitted through the monochromator.
Their height is less than 8\% compared to 400 eV peaks located closer to the central peak.
The randomly distributed black spots in Fig. \ref{fig:Figure3}(a) are due to contaminations formed at the surface of the detector at the end of the experiment. The streak visible in Figs. \ref{fig:Figure3} (b,d) is due to a shadow of the beamstop holder visible in Figs. \ref{fig:Figure3} (a,c).
\begin{figure}
        \centering
        \includegraphics[width = 1.0\columnwidth]{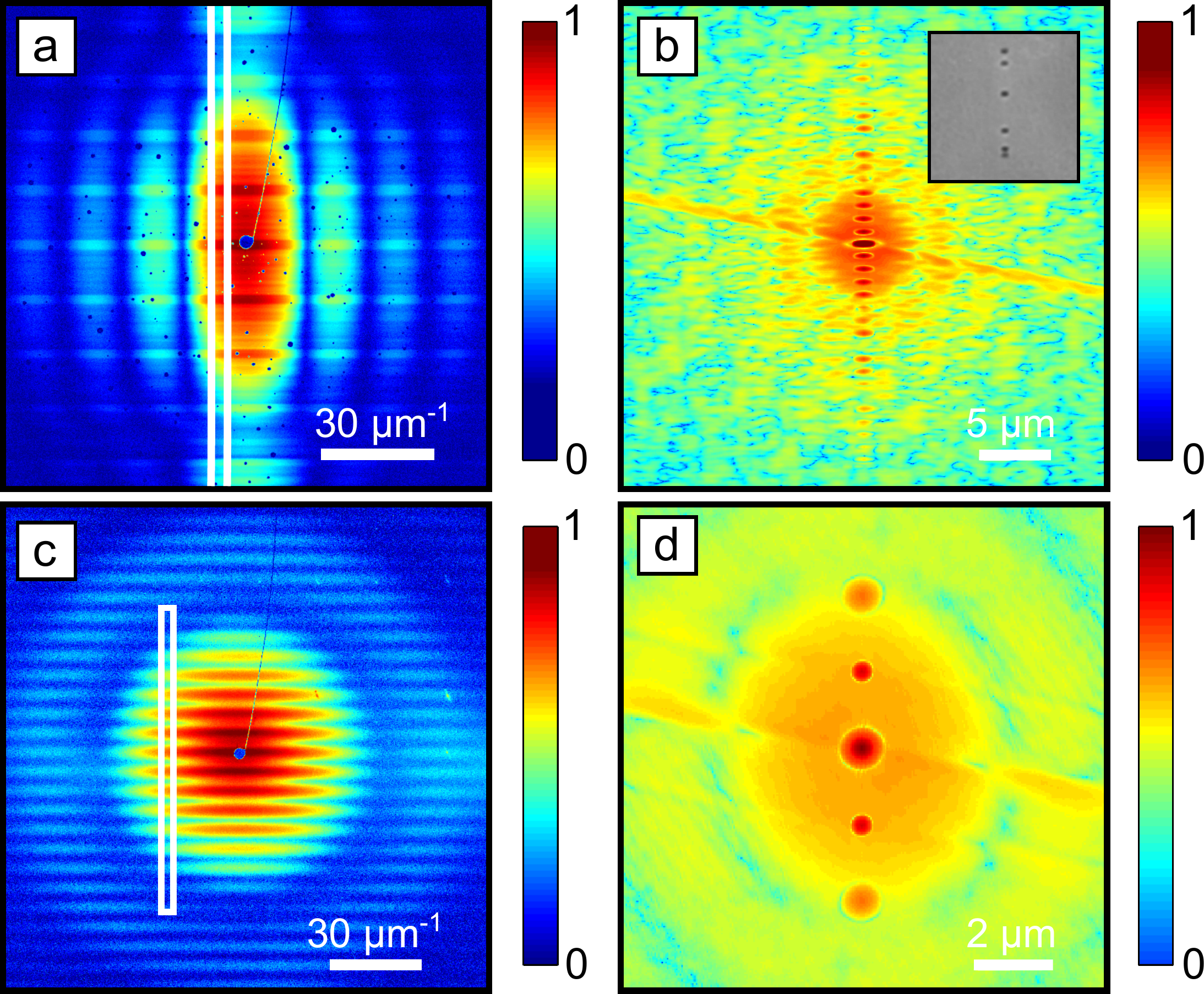}
    \caption{(a, c) Diffraction patterns from NRA apertures with six slits (a) and double pinholes with a separation of 2 $\mu$m (c)
    measured at 400 eV photon energy and monochromator exit slits size $D_\textrm{es}= 230$ $\mu$m.
    (b, d) Fourier transform of diffraction patterns in (a, c).
    All figures are displayed on a logarithmic scale.
    White rectangles in (a) and (c) indicate regions, which were used for the analysis.
    An optical microscope image of the NRA sample is shown in the inset of (b).
}
\label{fig:Figure3}
\end{figure}
The peak heights $C_0$ and $C_{ij}$ for both NRA and double pinholes diffraction patterns were retrieved from the Fourier transform of a vertical line scan, obtained by averaging 31 pixels wide regions marked by the white rectangle in Fig. \ref{fig:Figure3}(a) and Fig. \ref{fig:Figure3}(c).
Corresponding line scans of diffraction patterns and their Fourier Transforms are shown in Fig. \ref{fig:Figure4}.
In these regions high harmonic 800 eV radiation has a minimum contribution and temporal coherence effects can be neglected due to small scattering angles
\footnote{Our estimates show that the maximum optical path length difference $l$ is about $l=0.4$ $\mu$m  in the region chosen for the analysis.
This is smaller than the temporal coherence length $l_t=1.6$ $\mu$m and it means that our assumption $\tau < \tau_c$ is well satisfied in this region.}.
\begin{figure}
        \centering
        \includegraphics[width = 1.0\columnwidth]{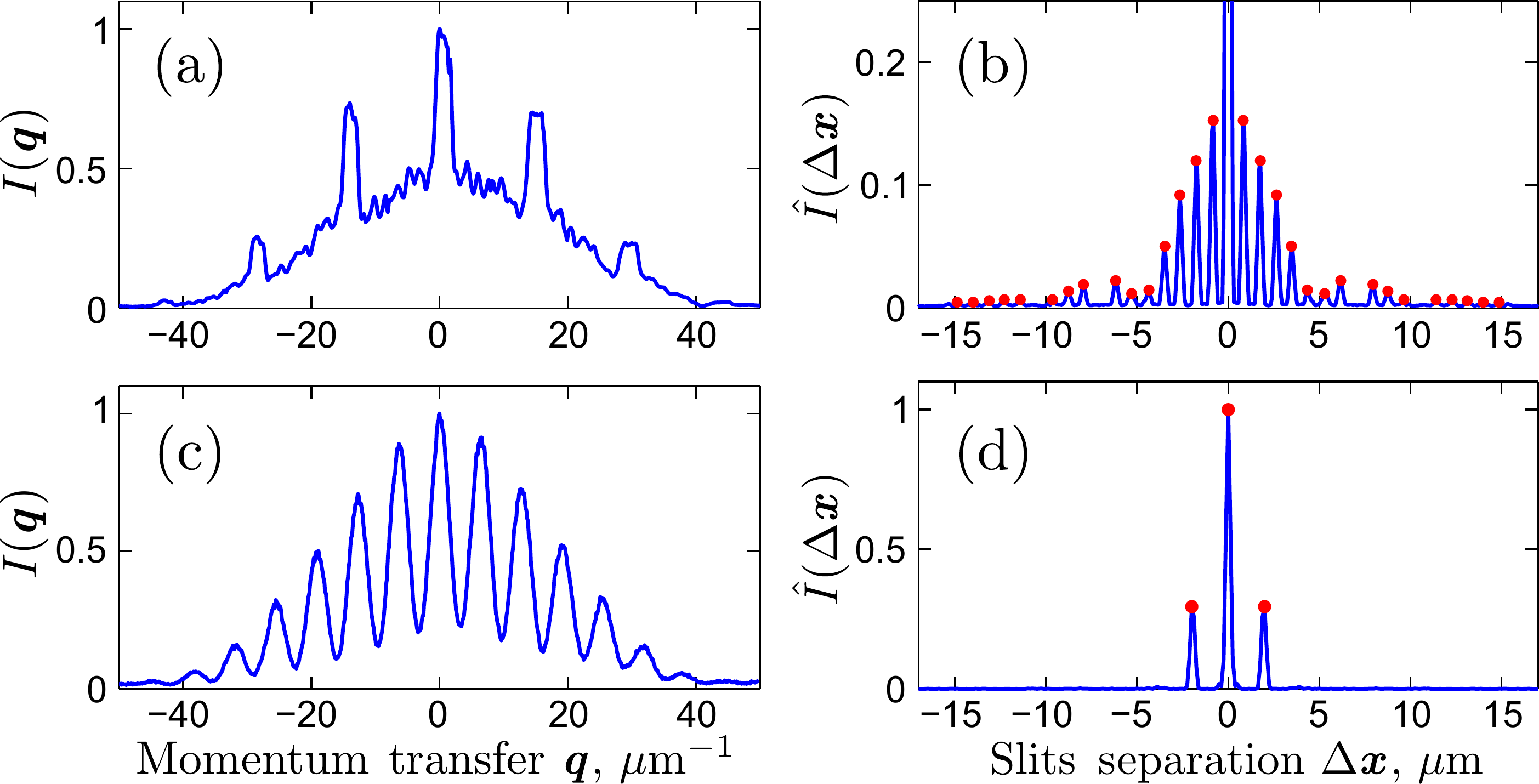}
    \caption{
    (a, c) Line scans of diffraction patterns $I(q)$ in the vertical direction from the NRA of six slits (a) and double pinholes (c) obtained by averaging the white regions shown in Figs. \ref{fig:Figure3}(a) and \ref{fig:Figure3}(b), respectively.
    (b, d) Fourier transforms $\hat I(\Delta x)$ of corresponding line scans from the NRA (b) and double pinholes (d).
     Red points in (b,d) denote the central and satellite peaks maxima $C_0$ and $C_{ij}$.}
    \label{fig:Figure4}
    \end{figure}
To determine the intensity distribution in the focus of the beam, the beam profile scans were performed by the double pinholes with a separation of 15 $\mu$m.
Each profile scan was fitted with two Gaussian functions with the same width and 15 $\mu$m separation.
The beam FWHM was determined to be in the range from 10 $\mu$m to 12 $\mu$m for the exit slits size $D_\textrm{es} = 40$ $\mu$m and in the range from 40 $\mu$m to 44 $\mu$m for $D_\textrm{es} = 230$ $\mu$m.
The beam size appeared to be independent of the photon energy and the beam defining slits width.
\begin{figure}
        \centering
        \includegraphics[width = 1.0\columnwidth]{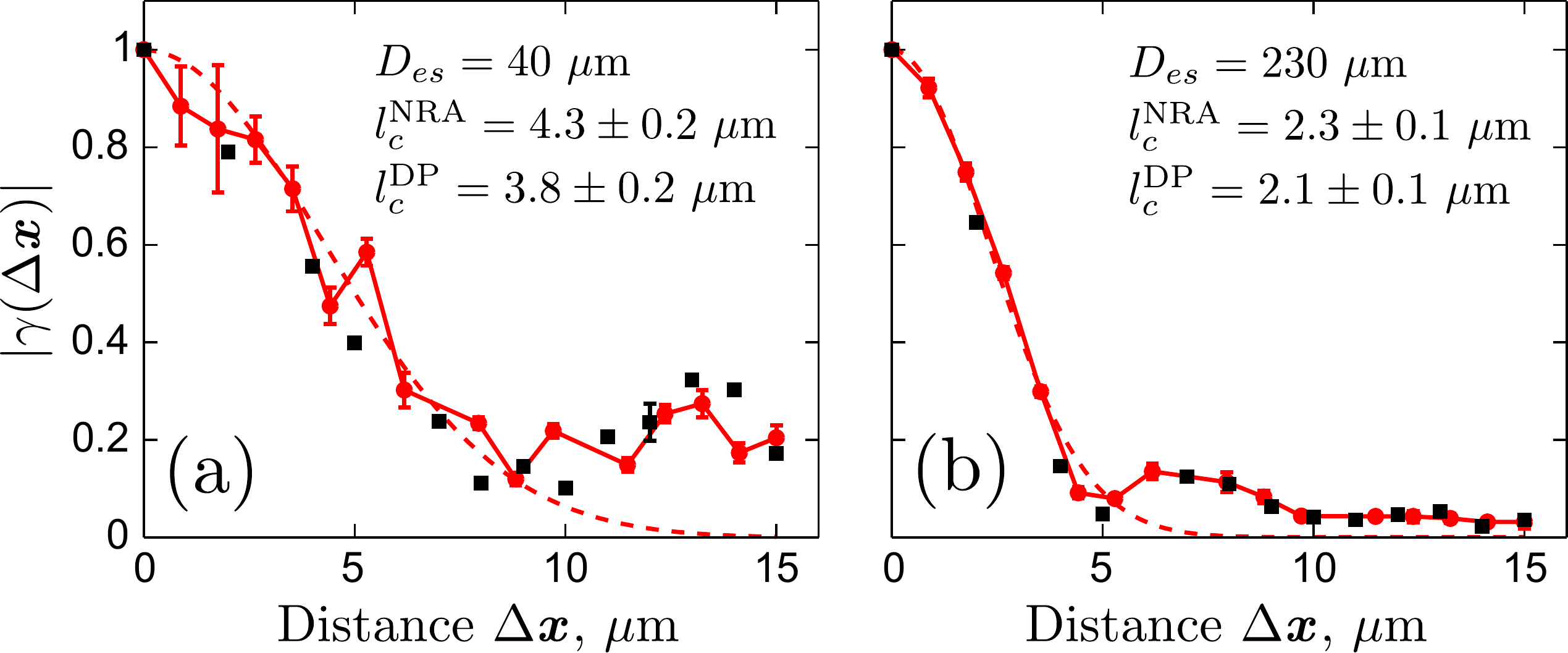}
    \caption{
     The modulus of the CCF $|\gamma(\Delta x)|$ for 400 eV synchrotron radiation and the monochromator exit slit width of $D_\textrm{es}= 40$ $\mu$m (a) and 230 $\mu$m (b).
     Red circles, connected with red lines, represent values of the CCF measured by NRA method and black squares correspond to the values determined by the double pinholes.  Gaussian fits for the CCF values obtained with NRAs are shown by dashed red lines.
            }
\label{fig:Figure5}
\end{figure}
Finally, the peak heights $C_{ij}$ and the measured incident intensity profiles $I_i$ were used in equation \eqref{eq:Eq4} to determine the modulus of the CCF $|\gamma_{ij}|$.
The result from the NRA measurements is presented in Fig. \ref{fig:Figure5} together with the double pinhole measurements.
It is important to note that the results obtained with the NRA of apertures are in excellent agreement with a well-established measurement using double pinholes, whereas the measurement time is reduced by an order of magnitude to acquire the same information.
\begin{table}[I]
\begin{tabular}{|c|c|c|}
\hline
 {\multirow{2}{*}{Parameter} } & \multicolumn{2}{c|}{Exit slit $D_\textrm{es}$, $\mu$m}  \\ \cline{2-3}
                {}             & 40               & 230                                               \\ \hline
  $l_c^\textrm{NRA}$, $\mu$m   & $4.3  \pm 0.2$   & $2.4  \pm 0.1$                                    \\ \cline{1-3}
  $l_c^\textrm{DP}$, $\mu$m    & $3.8  \pm 0.2$   & $2.1  \pm 0.1$                                    \\ \cline{1-3}
  $\zeta^\textrm{NRA}$         & $0.41 \pm 0.04$  & $0.06 \pm 0.01$                                   \\ \cline{1-3}
  $\zeta^\textrm{DP}$          & $0.38 \pm 0.03$  & $0.06 \pm 0.01$                                   \\ \hline
\end{tabular}\\
\caption{The coherence lengths $l_c^\textrm{NRA}$, $l_c^\textrm{DP}$ and the normalized degrees of coherence $\zeta^\textrm{NRA}$, $\zeta^\textrm{DP}$ measured with NRA and double pinholes (DP) for different exit slit widths of the monochromator $D_\textrm{es}$.
The beam had a FWHM of  $11\pm1$ $\mu$m in the case of $D_\textrm{es} = 40$ $\mu$m and $42\pm2$ $\mu$m in the case of $D_\textrm{es} = 230$ $\mu$m.
The photon energy was 400 eV and the beam defining slit was set to 4.7 mm.
}
\label{table:Table1}
\end{table}
The uncertainty of the CCF was estimated using a set of incident intensities within the error margins given by the intensity measurements.
Using equation \eqref{eq:Eq4} $|\gamma_{ij}|$ was calculated from each intensity profile and the uncertainty of $|\gamma_{ij}|$ was found from the deviation of these values.
The errors are comparably large when the beam size is small relative to the NRA size (that was 15 $\mu$m in our experiment), as in the case of 40 $\mu$m exit slit opening
\footnote{In fact, for the exit slit width of 40 $\mu$m the incident beam was smaller than the NRA size and we additionally had to find the position of the incident beam relative to the center of the NRA by putting additional constrains such as: avoiding values of $|\gamma_{ij}|$ larger than one, smallest uncertainties, and smoothness of the CCF.}.
When the FWHM of the focused beam was larger than the total NRA size, as in the case of 230 $\mu$m exit slit opening, practically no errors occurred (see Fig. \ref{fig:Figure5}(b)).
%From this analysis we can conclude that NRA method is especially well suited for the coherence measurements of the beams with the size  larger than the NRA.

To estimate the spatial coherence length $l_c$, the values of the modulus of the CCF $|\gamma(\Delta x)|$ were approximated by a Gaussian $\exp[-\Delta x^2/(2l^2_c)]$.
Results of our evaluation for the NRA as well as for the double pinholes are presented in Table I.
The values of the global degree of coherence $\zeta$ (see Eq. (3)) are also given in the same Table.
These results demonstrate an excellent agreement between two methods.
%As already seen in Fig. \ref{fig:Figure5} a good agreement is found between the NRA and double pinhole measurements, however, the performance of the NRA method is worse if the beam is too small.
%
%
\begin{figure}
        \centering
        \includegraphics[width = 1.0\columnwidth]{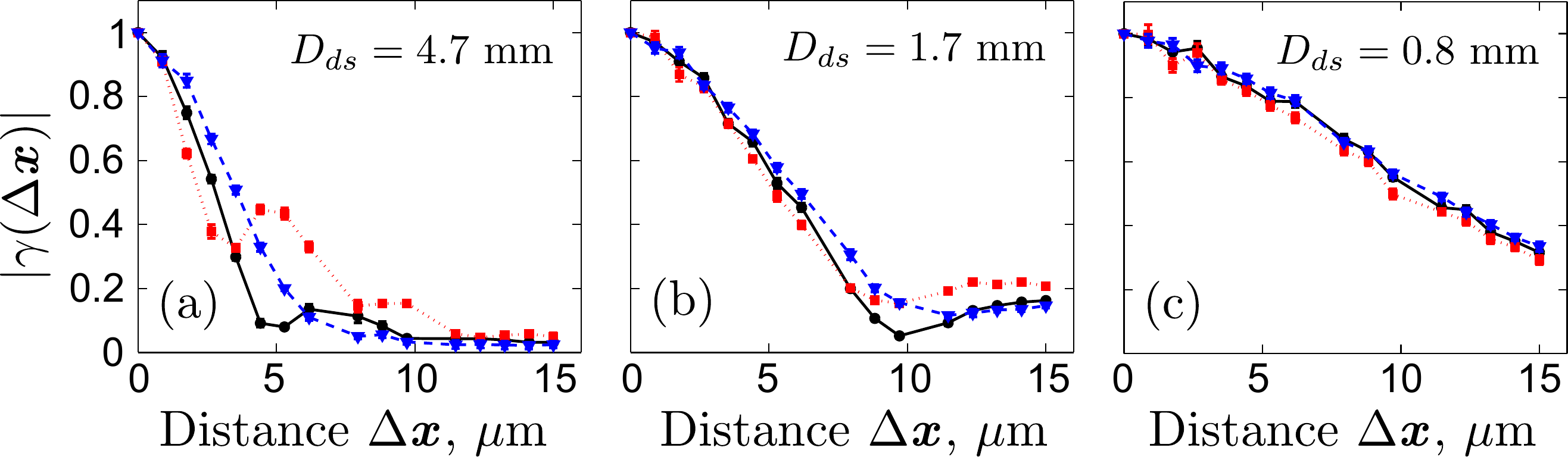}
          \caption{
           The modulus of the CCF $|\gamma(\Delta x)|$ measured for an exit slit width of $D_\textrm{es} = 230$ $\mu$m and beam defining slit widths of $D_\textrm{ds}=4.7$ mm (a), 1.7 mm (b), and 0.8 mm (c).
          Measurements performed at 396.5 eV (red squares), 400 eV (blue triangles)  and 403 eV (black circles) are presented.
          }
\label{fig:Figure6}
\end{figure}
Finally, we present spatial coherence measurements performed with NRA apertures for the cases of 396.5 eV, 400 eV, and 403 eV beam energies as well as 4.7 mm, 1.7 mm, and 0.8 mm beam defining slit width.
In all cases the monochromator exit slit width was set to $D_\textrm{es} = 230$ $\mu$m.
The results of these measurements are presented in Fig. \ref{fig:Figure6}.
The transverse coherence length $l_c$ for all curves was determined by the Gaussian fit, as before, and the result is summarized in Table \ref{table:Table2}.

As expected, $l_c$ is inversely proportional to the beam defining slit width.
The slit widths were chosen such, that the transmitted intensity drops by a factor of 2 in each step.
Interestingly, the normalized degree of coherence increases by about a factor 2 in each step.
This means that in the range accessed in our experiment the beam defining slit cuts out  the coherent part of the beam and almost no coherent flux is lost.
For the blue shifted beam (403 eV) this relation does not hold due to a high value of the normalized degree of coherence \cite{SingerSPIE2011}.

We also observed a variation of the transverse coherence length as a function of the monochromator tuning energy.
The energies were chosen such that the maximum flux at 400 eV was reduced by a factor of 2 for both energies 403 eV and 396.5 eV.
For the largest beam defining slit (4.7 mm) the coherence length increases by 25 \% when the monochromator is detuned to 403 eV.
However, unlike in the case of the beam defining slit the coherent flux is reduced.
Interestingly, the functional form appears to be closer to a Gaussian function for the blue shifted radiation.
We want to note here, that it is the power of the method here that allows to record these fine details of the coherence function.
For an energy of 396.5 eV the transverse coherence is worse \footnote{This effect could be due to the fact that on the higher energy side of the undulator peak the collimation of the undulator cone is better than towards the lower energy side where at even lower energies an intensity minimum is observed on axis, resulting in a non-optimal beam path.}.
The energy dependence of the transverse coherence is significantly smaller (about 5\%) for the smallest beam defining slit.
%Also, absolute CCF oscillations are strongly pronounced for the widest opening of the beam defining slit (see Fig.\ref{fig:Figure5}.(a)). The oscillations could have arisen due to wavefield diffraction on the edges of the exit slits. A similar effect was observed in \cite{AlaimoPRL2009}.

\begin{table}[II]
%\begin{tabular}{|C{17mm}|C{26mm}|C{34mm}|C{34mm}|C{34mm}|}
\begin{tabular}{|c|c|c|c|c|}
\hline
 {\multirow{2}{*}{Energy, eV}  } & {\multirow{2}{*}{Parameter} }  & \multicolumn{3}{c|}{Beam defining slit $D_\textrm{ds}$, mm}       \\ \cline{3-5}
 {}                          &               {}               & 4.7            & 1.7              &    0.8           \\ \hline
 {\multirow{2}{*}{396.5}} & $l_c$, $\mu$m                  & $1.9  \pm 0.1$   & $4.4  \pm 0.1$     &    $8.8  \pm 0.3$   \\ \cline{2-5}
 {}                          & $\zeta$                        & $0.05 \pm 0.01$  & $0.12 \pm 0.01$    &    $0.24 \pm 0.01$  \\ \hline
 {\multirow{2}{*}{400}}   & $l_c$, $\mu$m                  & $2.4  \pm 0.1$   & $4.6  \pm 0.1$     &    $9.2  \pm 0.2$   \\ \cline{2-5}
 {}                          & $\zeta$                        & $0.06 \pm 0.01$  & $0.13 \pm 0.01$    &    $0.25 \pm 0.01$  \\ \hline
 {\multirow{2}{*}{403}}   & $l_c$, $\mu$m                  & $3.0  \pm 0.1$   & $5.1  \pm 0.1$     &    $9.4  \pm 0.2$   \\ \cline{2-5}
 {}                          & $\zeta$                        & $0.08 \pm 0.01$  & $0.14 \pm 0.01$    &    $0.25 \pm 0.01$  \\ \hline
\end{tabular}\\
\caption{The coherence length $l_c$ and normalized degree of coherence $\zeta$ for different photon energies and the beam defining slit size $D_\textrm{ds}$ obtained as a result of the analysis of data presented in Fig. 6.
The monochromator exit slit were fixed to $D_\textrm{es}=230$ $\mu$m.}
\label{table:Table2}
\end{table}

    \section{Conclusions and Outlook}

In conclusion, a non-redundant array of apertures provides a fast and effective way to fully determine the spatial coherence function of the undulator radiation.
We showed that in this method a single diffraction pattern is sufficient to find degree of coherence of X-ray radiation at several relative distances simultaneously.
The results obtained with NRA apertures concord well with the values determined using Young's double pinholes, whereas the measurement time was reduced significantly.
The dependence of spatial coherence on different parameters was explored.
The transverse coherence length shows an inverse proportional dependence on both the width of the beam defining slit and the exit slit.
We also observed that the spatial coherence improves if the monochromator is offset to larger energies (blue shifted) and it is decreased for smaller energies (red shifted).
We anticipate that the method presented here will be a useful procedure for XUV and soft X-ray beam calibration prior to any experiment utilizing spatial coherence.
We also foresee that in future the NRA method could become a viable tool for single pulse coherence characterization measurements at free-electron lasers \cite{Singer_FLASH_2008,Vartanyants2010,VartanyantsPRL2011,Singer_FLASH_2012}.

    \section{Acknowledgements}
We acknowledge fruitful discussions and support of the project by E. Weckert, help with the manufacturing of the NRA apertures by D. Attwood, and help with the design of the NRA apertures by A. Mancuso. We are thankful to F. Scholz, I. Shevchuk and J. Seltmann for setting up the experiment at P04 beamline. Part of this work was supported by BMBF Proposal 05K10CHG `'Coherent Diffraction Imaging and Scattering of Ultrashort Coherent Pulses with Matter`' in the framework of the German-Russian collaboration `'Development and Use of Accelerator-Based Photon Sources`' and the Virtual Institute VH-VI-403 of the Helmholtz Association. Additional support by BMBF under 05K10GU4 and from the Hamburg Centre for Ultrafast Imaging is greatly acknowledged.

%\referencelist

\bibliography{P04_Bibliography}

%merlin.mbs apsrev4-1.bst 2010-07-25 4.21a (PWD, AO, DPC) hacked
%Control: key (0)
%Control: author (8) initials jnrlst
%Control: editor formatted (1) identically to author
%Control: production of article title (-1) disabled
%Control: page (0) single
%Control: year (1) truncated
%Control: production of eprint (0) enabled
\begin{thebibliography}{42}%
\makeatletter
\providecommand \@ifxundefined [1]{%
 \@ifx{#1\undefined}
}%
\providecommand \@ifnum [1]{%
 \ifnum #1\expandafter \@firstoftwo
 \else \expandafter \@secondoftwo
 \fi
}%
\providecommand \@ifx [1]{%
 \ifx #1\expandafter \@firstoftwo
 \else \expandafter \@secondoftwo
 \fi
}%
\providecommand \natexlab [1]{#1}%
\providecommand \enquote  [1]{``#1''}%
\providecommand \bibnamefont  [1]{#1}%
\providecommand \bibfnamefont [1]{#1}%
\providecommand \citenamefont [1]{#1}%
\providecommand \href@noop [0]{\@secondoftwo}%
\providecommand \href [0]{\begingroup \@sanitize@url \@href}%
\providecommand \@href[1]{\@@startlink{#1}\@@href}%
\providecommand \@@href[1]{\endgroup#1\@@endlink}%
\providecommand \@sanitize@url [0]{\catcode `\\12\catcode `\$12\catcode
  `\&12\catcode `\#12\catcode `\^12\catcode `\_12\catcode `\%12\relax}%
\providecommand \@@startlink[1]{}%
\providecommand \@@endlink[0]{}%
\providecommand \url  [0]{\begingroup\@sanitize@url \@url }%
\providecommand \@url [1]{\endgroup\@href {#1}{\urlprefix }}%
\providecommand \urlprefix  [0]{URL }%
\providecommand \Eprint [0]{\href }%
\providecommand \doibase [0]{http://dx.doi.org/}%
\providecommand \selectlanguage [0]{\@gobble}%
\providecommand \bibinfo  [0]{\@secondoftwo}%
\providecommand \bibfield  [0]{\@secondoftwo}%
\providecommand \translation [1]{[#1]}%
\providecommand \BibitemOpen [0]{}%
\providecommand \bibitemStop [0]{}%
\providecommand \bibitemNoStop [0]{.\EOS\space}%
\providecommand \EOS [0]{\spacefactor3000\relax}%
\providecommand \BibitemShut  [1]{\csname bibitem#1\endcsname}%
\let\auto@bib@innerbib\@empty
%</preamble>
\bibitem [{\citenamefont {Winick}(1980)}]{Winick}%
  \BibitemOpen
  \bibfield  {author} {\bibinfo {author} {\bibfnamefont {H.}~\bibnamefont
  {Winick}},\ }\href@noop {} {\emph {\bibinfo {title} {Synchrotron radiation
  research: Edited by H Winick and S Doniach.}}}\ (\bibinfo  {publisher}
  {Plenum},\ \bibinfo {address} {New York and London},\ \bibinfo {year}
  {1980})\BibitemShut {NoStop}%
\bibitem [{\citenamefont {Attwood}(1999)}]{Attwood}%
  \BibitemOpen
  \bibfield  {author} {\bibinfo {author} {\bibfnamefont {D.}~\bibnamefont
  {Attwood}},\ }\href@noop {} {\emph {\bibinfo {title} {Soft X-rays and Extreme
  Ultraviolet Radiation}}}\ (\bibinfo  {publisher} {Cambridge Univ. Press},\
  \bibinfo {address} {Cambridge},\ \bibinfo {year} {1999})\BibitemShut
  {NoStop}%
\bibitem [{\citenamefont {Vartanyants}\ and\ \citenamefont
  {Singer}(2010)}]{VartanyantsNJP2010}%
  \BibitemOpen
  \bibfield  {author} {\bibinfo {author} {\bibfnamefont {I.~A.}\ \bibnamefont
  {Vartanyants}}\ and\ \bibinfo {author} {\bibfnamefont {A.}~\bibnamefont
  {Singer}},\ }\href {http://stacks.iop.org/1367-2630/12/i=3/a=035004}
  {\bibfield  {journal} {\bibinfo  {journal} {New J. Phys.}\ }\textbf {\bibinfo
  {volume} {12}},\ \bibinfo {pages} {035004} (\bibinfo {year}
  {2010})}\BibitemShut {NoStop}%
\bibitem [{\citenamefont {Vartanyants}\ \emph {et~al.}(2010)\citenamefont
  {Vartanyants}, \citenamefont {Mancuso}, \citenamefont {Singer}, \citenamefont
  {Yefanov},\ and\ \citenamefont {Gulden}}]{Vartanyants2010}%
  \BibitemOpen
  \bibfield  {author} {\bibinfo {author} {\bibfnamefont {I.~A.}\ \bibnamefont
  {Vartanyants}}, \bibinfo {author} {\bibfnamefont {A.~P.}\ \bibnamefont
  {Mancuso}}, \bibinfo {author} {\bibfnamefont {A.}~\bibnamefont {Singer}},
  \bibinfo {author} {\bibfnamefont {O.~M.}\ \bibnamefont {Yefanov}}, \ and\
  \bibinfo {author} {\bibfnamefont {J.}~\bibnamefont {Gulden}},\ }\href
  {http://stacks.iop.org/0953-4075/43/i=19/a=194016} {\bibfield  {journal}
  {\bibinfo  {journal} {J. Phys. B}\ }\textbf {\bibinfo {volume} {43}},\
  \bibinfo {pages} {194016} (\bibinfo {year} {2010})}\BibitemShut {NoStop}%
\bibitem [{\citenamefont {Chapman}\ and\ \citenamefont
  {Nugent}(2010)}]{ChapmanNatPhot2010}%
  \BibitemOpen
  \bibfield  {author} {\bibinfo {author} {\bibfnamefont {H.~N.}\ \bibnamefont
  {Chapman}}\ and\ \bibinfo {author} {\bibfnamefont {K.~A.}\ \bibnamefont
  {Nugent}},\ }\href {\doibase 10.1038/nphoton.2010.240} {\bibfield  {journal}
  {\bibinfo  {journal} {Nat. Photon}\ }\textbf {\bibinfo {volume} {4}},\
  \bibinfo {pages} {833} (\bibinfo {year} {2010})}\BibitemShut {NoStop}%
\bibitem [{\citenamefont {Mancuso}\ \emph {et~al.}(2010)\citenamefont
  {Mancuso}, \citenamefont {Gorniak}, \citenamefont {Staier}, \citenamefont
  {Yefanov}, \citenamefont {Barth}, \citenamefont {Christophis}, \citenamefont
  {Reime}, \citenamefont {Gulden}, \citenamefont {Singer}, \citenamefont
  {Pettit}, \citenamefont {Nisius}, \citenamefont {Wilhein}, \citenamefont
  {Gutt}, \citenamefont {Gr{\"u}bel}, \citenamefont {Guerassimova},
  \citenamefont {Treusch}, \citenamefont {Feldhaus}, \citenamefont {Eisebitt},
  \citenamefont {Weckert}, \citenamefont {Grunze}, \citenamefont {Rosenhahn},\
  and\ \citenamefont {Vartanyants}}]{MancusoNJP2010}%
  \BibitemOpen
  \bibfield  {author} {\bibinfo {author} {\bibfnamefont {A.~P.}\ \bibnamefont
  {Mancuso}}, \bibinfo {author} {\bibfnamefont {T.}~\bibnamefont {Gorniak}},
  \bibinfo {author} {\bibfnamefont {F.}~\bibnamefont {Staier}}, \bibinfo
  {author} {\bibfnamefont {O.~M.}\ \bibnamefont {Yefanov}}, \bibinfo {author}
  {\bibfnamefont {R.}~\bibnamefont {Barth}}, \bibinfo {author} {\bibfnamefont
  {C.}~\bibnamefont {Christophis}}, \bibinfo {author} {\bibfnamefont
  {B.}~\bibnamefont {Reime}}, \bibinfo {author} {\bibfnamefont
  {J.}~\bibnamefont {Gulden}}, \bibinfo {author} {\bibfnamefont
  {A.}~\bibnamefont {Singer}}, \bibinfo {author} {\bibfnamefont {M.~E.}\
  \bibnamefont {Pettit}}, \bibinfo {author} {\bibfnamefont {T.}~\bibnamefont
  {Nisius}}, \bibinfo {author} {\bibfnamefont {T.}~\bibnamefont {Wilhein}},
  \bibinfo {author} {\bibfnamefont {C.}~\bibnamefont {Gutt}}, \bibinfo {author}
  {\bibfnamefont {G.}~\bibnamefont {Gr{\"u}bel}}, \bibinfo {author}
  {\bibfnamefont {N.}~\bibnamefont {Guerassimova}}, \bibinfo {author}
  {\bibfnamefont {R.}~\bibnamefont {Treusch}}, \bibinfo {author} {\bibfnamefont
  {J.}~\bibnamefont {Feldhaus}}, \bibinfo {author} {\bibfnamefont
  {S.}~\bibnamefont {Eisebitt}}, \bibinfo {author} {\bibfnamefont
  {E.}~\bibnamefont {Weckert}}, \bibinfo {author} {\bibfnamefont
  {M.}~\bibnamefont {Grunze}}, \bibinfo {author} {\bibfnamefont
  {A.}~\bibnamefont {Rosenhahn}}, \ and\ \bibinfo {author} {\bibfnamefont
  {I.~A.}\ \bibnamefont {Vartanyants}},\ }\href
  {http://stacks.iop.org/1367-2630/12/i=3/a=035003} {\bibfield  {journal}
  {\bibinfo  {journal} {New J. Phys.}\ }\textbf {\bibinfo {volume} {12}},\
  \bibinfo {pages} {035003} (\bibinfo {year} {2010})}\BibitemShut {NoStop}%
\bibitem [{\citenamefont {Eisebitt}\ \emph {et~al.}(2010)\citenamefont
  {Eisebitt}, \citenamefont {Lüning}, \citenamefont {Schlotter}, \citenamefont
  {Lörgen}, \citenamefont {Hellwig}, \citenamefont {Eberhardt},\ and\
  \citenamefont {Stöhr}}]{Eisebitt_Nature_2004}%
  \BibitemOpen
  \bibfield  {author} {\bibinfo {author} {\bibfnamefont {S.}~\bibnamefont
  {Eisebitt}}, \bibinfo {author} {\bibfnamefont {J.}~\bibnamefont {Lüning}},
  \bibinfo {author} {\bibfnamefont {W.~F.}\ \bibnamefont {Schlotter}}, \bibinfo
  {author} {\bibfnamefont {M.}~\bibnamefont {Lörgen}}, \bibinfo {author}
  {\bibfnamefont {O.}~\bibnamefont {Hellwig}}, \bibinfo {author} {\bibfnamefont
  {W.}~\bibnamefont {Eberhardt}}, \ and\ \bibinfo {author} {\bibfnamefont
  {J.}~\bibnamefont {Stöhr}},\ }\href
  {http://scitation.aip.org/content/aip/journal/apl/96/4/10.1063/1.3291942}
  {\bibfield  {journal} {\bibinfo  {journal} {Nature}\ }\textbf {\bibinfo
  {volume} {95}},\ \bibinfo {pages} {042501} (\bibinfo {year}
  {2010})}\BibitemShut {NoStop}%
\bibitem [{\citenamefont {Stickler1}\ \emph {et~al.}(2004)\citenamefont
  {Stickler1}, \citenamefont {Fr\"{o}mter}, \citenamefont {Stillrich},
  \citenamefont {Menk}, \citenamefont {Tieg}, \citenamefont
  {Streit-Nierobisch}, \citenamefont {Sprung}, \citenamefont {Gutt},
  \citenamefont {Stadler}, \citenamefont {Leupold}, \citenamefont {Grübel},\
  and\ \citenamefont {Oepen}}]{Stickler_APL_2010}%
  \BibitemOpen
  \bibfield  {author} {\bibinfo {author} {\bibfnamefont {D.}~\bibnamefont
  {Stickler1}}, \bibinfo {author} {\bibfnamefont {R.}~\bibnamefont
  {Fr\"{o}mter}}, \bibinfo {author} {\bibfnamefont {H.}~\bibnamefont
  {Stillrich}}, \bibinfo {author} {\bibfnamefont {C.}~\bibnamefont {Menk}},
  \bibinfo {author} {\bibfnamefont {C.}~\bibnamefont {Tieg}}, \bibinfo {author}
  {\bibfnamefont {S.}~\bibnamefont {Streit-Nierobisch}}, \bibinfo {author}
  {\bibfnamefont {M.}~\bibnamefont {Sprung}}, \bibinfo {author} {\bibfnamefont
  {C.}~\bibnamefont {Gutt}}, \bibinfo {author} {\bibfnamefont {L.-M.}\
  \bibnamefont {Stadler}}, \bibinfo {author} {\bibfnamefont {O.}~\bibnamefont
  {Leupold}}, \bibinfo {author} {\bibfnamefont {G.}~\bibnamefont {Grübel}}, \
  and\ \bibinfo {author} {\bibfnamefont {H.~P.}\ \bibnamefont {Oepen}},\ }\href
  {http://www.nature.com/nature/journal/v432/n7019/abs/nature03139.html}
  {\bibfield  {journal} {\bibinfo  {journal} {Applied Physics Letters}\
  }\textbf {\bibinfo {volume} {432}},\ \bibinfo {pages} {885} (\bibinfo {year}
  {2004})}\BibitemShut {NoStop}%
\bibitem [{\citenamefont {Gr{\"u}bel}\ and\ \citenamefont
  {Zontone}(2004)}]{GruebelJAC2004}%
  \BibitemOpen
  \bibfield  {author} {\bibinfo {author} {\bibfnamefont {G.}~\bibnamefont
  {Gr{\"u}bel}}\ and\ \bibinfo {author} {\bibfnamefont {F.}~\bibnamefont
  {Zontone}},\ }\href {\doibase 10.1016/S0925-8388(03)00555-3} {\bibfield
  {journal} {\bibinfo  {journal} {J. Alloys Compd.}\ }\textbf {\bibinfo
  {volume} {362}},\ \bibinfo {pages} {3 } (\bibinfo {year} {2004})}\BibitemShut
  {NoStop}%
\bibitem [{\citenamefont {Fienup}(1982)}]{FienupApplOpt1982}%
  \BibitemOpen
  \bibfield  {author} {\bibinfo {author} {\bibfnamefont {J.~R.}\ \bibnamefont
  {Fienup}},\ }\href {\doibase 10.1364/AO.21.002758} {\bibfield  {journal}
  {\bibinfo  {journal} {Appl. Opt.}\ }\textbf {\bibinfo {volume} {21}},\
  \bibinfo {pages} {2758} (\bibinfo {year} {1982})}\BibitemShut {NoStop}%
\bibitem [{\citenamefont {Goodman}(2007)}]{G2007}%
  \BibitemOpen
  \bibfield  {author} {\bibinfo {author} {\bibfnamefont {J.~W.}\ \bibnamefont
  {Goodman}},\ }\href@noop {} {\emph {\bibinfo {title} {Speckle Phenomena in
  Optics}}}\ (\bibinfo  {publisher} {Roberts and Company Publishers},\ \bibinfo
  {address} {Greenwood Village},\ \bibinfo {year} {2007})\BibitemShut {NoStop}%
\bibitem [{\citenamefont {Whitehead}\ \emph {et~al.}(2009)\citenamefont
  {Whitehead}, \citenamefont {Williams}, \citenamefont {Quiney}, \citenamefont
  {Vine}, \citenamefont {Dilanian}, \citenamefont {Flewett}, \citenamefont
  {Nugent}, \citenamefont {Peele}, \citenamefont {Balaur},\ and\ \citenamefont
  {McNulty}}]{WhiteheadPRL2009}%
  \BibitemOpen
  \bibfield  {author} {\bibinfo {author} {\bibfnamefont {L.~W.}\ \bibnamefont
  {Whitehead}}, \bibinfo {author} {\bibfnamefont {G.~J.}\ \bibnamefont
  {Williams}}, \bibinfo {author} {\bibfnamefont {H.~M.}\ \bibnamefont
  {Quiney}}, \bibinfo {author} {\bibfnamefont {D.~J.}\ \bibnamefont {Vine}},
  \bibinfo {author} {\bibfnamefont {R.~A.}\ \bibnamefont {Dilanian}}, \bibinfo
  {author} {\bibfnamefont {S.}~\bibnamefont {Flewett}}, \bibinfo {author}
  {\bibfnamefont {K.~A.}\ \bibnamefont {Nugent}}, \bibinfo {author}
  {\bibfnamefont {A.~G.}\ \bibnamefont {Peele}}, \bibinfo {author}
  {\bibfnamefont {E.}~\bibnamefont {Balaur}}, \ and\ \bibinfo {author}
  {\bibfnamefont {I.}~\bibnamefont {McNulty}},\ }\href {\doibase
  10.1103/PhysRevLett.103.243902} {\bibfield  {journal} {\bibinfo  {journal}
  {Phys. Rev. Lett.}\ }\textbf {\bibinfo {volume} {103}},\ \bibinfo {pages}
  {243902} (\bibinfo {year} {2009})}\BibitemShut {NoStop}%
\bibitem [{\citenamefont {Chang}\ \emph {et~al.}(2000)\citenamefont {Chang},
  \citenamefont {Naulleau}, \citenamefont {Anderson},\ and\ \citenamefont
  {Attwood}}]{ChangOptCommun2000}%
  \BibitemOpen
  \bibfield  {author} {\bibinfo {author} {\bibfnamefont {C.}~\bibnamefont
  {Chang}}, \bibinfo {author} {\bibfnamefont {P.}~\bibnamefont {Naulleau}},
  \bibinfo {author} {\bibfnamefont {E.}~\bibnamefont {Anderson}}, \ and\
  \bibinfo {author} {\bibfnamefont {D.}~\bibnamefont {Attwood}},\ }\href
  {\doibase 10.1016/S0030-4018(00)00811-7} {\bibfield  {journal} {\bibinfo
  {journal} {Opt. Commun.}\ }\textbf {\bibinfo {volume} {182}},\ \bibinfo
  {pages} {25 } (\bibinfo {year} {2000})}\BibitemShut {NoStop}%
\bibitem [{\citenamefont {{Paterson}}\ \emph {et~al.}(2001)\citenamefont
  {{Paterson}}, \citenamefont {{Allman}}, \citenamefont {{McMahon}},
  \citenamefont {{Lin}}, \citenamefont {{Moldovan}}, \citenamefont {{Nugent}},
  \citenamefont {{McNulty}}, \citenamefont {{Chantler}}, \citenamefont
  {{Retsch}}, \citenamefont {{Irving}},\ and\ \citenamefont
  {{Mancini}}}]{NugentSynchr2001}%
  \BibitemOpen
  \bibfield  {author} {\bibinfo {author} {\bibfnamefont {D.}~\bibnamefont
  {{Paterson}}}, \bibinfo {author} {\bibfnamefont {B.~E.}\ \bibnamefont
  {{Allman}}}, \bibinfo {author} {\bibfnamefont {P.~J.}\ \bibnamefont
  {{McMahon}}}, \bibinfo {author} {\bibfnamefont {J.}~\bibnamefont {{Lin}}},
  \bibinfo {author} {\bibfnamefont {N.}~\bibnamefont {{Moldovan}}}, \bibinfo
  {author} {\bibfnamefont {K.~A.}\ \bibnamefont {{Nugent}}}, \bibinfo {author}
  {\bibfnamefont {I.}~\bibnamefont {{McNulty}}}, \bibinfo {author}
  {\bibfnamefont {C.~T.}\ \bibnamefont {{Chantler}}}, \bibinfo {author}
  {\bibfnamefont {C.~C.}\ \bibnamefont {{Retsch}}}, \bibinfo {author}
  {\bibfnamefont {T.~H.~K.}\ \bibnamefont {{Irving}}}, \ and\ \bibinfo {author}
  {\bibfnamefont {D.~C.}\ \bibnamefont {{Mancini}}},\ }\href {\doibase
  10.1016/S0030-4018(01)01276-7} {\bibfield  {journal} {\bibinfo  {journal}
  {Opt. Commun.}\ }\textbf {\bibinfo {volume} {195}},\ \bibinfo {pages} {79}
  (\bibinfo {year} {2001})}\BibitemShut {NoStop}%
\bibitem [{\citenamefont {Singer}\ \emph {et~al.}(2008)\citenamefont {Singer},
  \citenamefont {Vartanyants}, \citenamefont {Kuhlmann}, \citenamefont
  {Duesterer}, \citenamefont {Treusch},\ and\ \citenamefont
  {Feldhaus}}]{Singer_FLASH_2008}%
  \BibitemOpen
  \bibfield  {author} {\bibinfo {author} {\bibfnamefont {A.}~\bibnamefont
  {Singer}}, \bibinfo {author} {\bibfnamefont {I.~A.}\ \bibnamefont
  {Vartanyants}}, \bibinfo {author} {\bibfnamefont {M.}~\bibnamefont
  {Kuhlmann}}, \bibinfo {author} {\bibfnamefont {S.}~\bibnamefont {Duesterer}},
  \bibinfo {author} {\bibfnamefont {R.}~\bibnamefont {Treusch}}, \ and\
  \bibinfo {author} {\bibfnamefont {J.}~\bibnamefont {Feldhaus}},\ }\href
  {\doibase 10.1103/PhysRevLett.101.254801} {\bibfield  {journal} {\bibinfo
  {journal} {Phys. Rev. Lett.}\ }\textbf {\bibinfo {volume} {101}},\ \bibinfo
  {pages} {254801} (\bibinfo {year} {2008})}\BibitemShut {NoStop}%
\bibitem [{\citenamefont {Vartanyants}\ \emph
  {et~al.}(2011{\natexlab{a}})\citenamefont {Vartanyants}, \citenamefont
  {Singer}, \citenamefont {Mancuso}, \citenamefont {Yefanov}, \citenamefont
  {Sakdinawat}, \citenamefont {Liu}, \citenamefont {Bang}, \citenamefont
  {Williams}, \citenamefont {Cadenazzi}, \citenamefont {Abbey}, \citenamefont
  {Sinn}, \citenamefont {Attwood}, \citenamefont {Nugent}, \citenamefont
  {Weckert}, \citenamefont {Wang}, \citenamefont {Zhu}, \citenamefont {Wu},
  \citenamefont {Graves}, \citenamefont {Scherz}, \citenamefont {Turner},
  \citenamefont {Schlotter}, \citenamefont {Messerschmidt}, \citenamefont
  {L\"uning}, \citenamefont {Acremann}, \citenamefont {Heimann}, \citenamefont
  {Mancini}, \citenamefont {Joshi}, \citenamefont {Krzywinski}, \citenamefont
  {Soufli}, \citenamefont {Fernandez-Perea}, \citenamefont {Hau-Riege},
  \citenamefont {Peele}, \citenamefont {Feng}, \citenamefont {Krupin},
  \citenamefont {Moeller},\ and\ \citenamefont {Wurth}}]{Singer_LCLS_2011}%
  \BibitemOpen
  \bibfield  {author} {\bibinfo {author} {\bibfnamefont {I.~A.}\ \bibnamefont
  {Vartanyants}}, \bibinfo {author} {\bibfnamefont {A.}~\bibnamefont {Singer}},
  \bibinfo {author} {\bibfnamefont {A.~P.}\ \bibnamefont {Mancuso}}, \bibinfo
  {author} {\bibfnamefont {O.~M.}\ \bibnamefont {Yefanov}}, \bibinfo {author}
  {\bibfnamefont {A.}~\bibnamefont {Sakdinawat}}, \bibinfo {author}
  {\bibfnamefont {Y.}~\bibnamefont {Liu}}, \bibinfo {author} {\bibfnamefont
  {E.}~\bibnamefont {Bang}}, \bibinfo {author} {\bibfnamefont {G.~J.}\
  \bibnamefont {Williams}}, \bibinfo {author} {\bibfnamefont {G.}~\bibnamefont
  {Cadenazzi}}, \bibinfo {author} {\bibfnamefont {B.}~\bibnamefont {Abbey}},
  \bibinfo {author} {\bibfnamefont {H.}~\bibnamefont {Sinn}}, \bibinfo {author}
  {\bibfnamefont {D.}~\bibnamefont {Attwood}}, \bibinfo {author} {\bibfnamefont
  {K.~A.}\ \bibnamefont {Nugent}}, \bibinfo {author} {\bibfnamefont
  {E.}~\bibnamefont {Weckert}}, \bibinfo {author} {\bibfnamefont
  {T.}~\bibnamefont {Wang}}, \bibinfo {author} {\bibfnamefont {D.}~\bibnamefont
  {Zhu}}, \bibinfo {author} {\bibfnamefont {B.}~\bibnamefont {Wu}}, \bibinfo
  {author} {\bibfnamefont {C.}~\bibnamefont {Graves}}, \bibinfo {author}
  {\bibfnamefont {A.}~\bibnamefont {Scherz}}, \bibinfo {author} {\bibfnamefont
  {J.~J.}\ \bibnamefont {Turner}}, \bibinfo {author} {\bibfnamefont {W.~F.}\
  \bibnamefont {Schlotter}}, \bibinfo {author} {\bibfnamefont {M.}~\bibnamefont
  {Messerschmidt}}, \bibinfo {author} {\bibfnamefont {J.}~\bibnamefont
  {L\"uning}}, \bibinfo {author} {\bibfnamefont {Y.}~\bibnamefont {Acremann}},
  \bibinfo {author} {\bibfnamefont {P.}~\bibnamefont {Heimann}}, \bibinfo
  {author} {\bibfnamefont {D.~C.}\ \bibnamefont {Mancini}}, \bibinfo {author}
  {\bibfnamefont {V.}~\bibnamefont {Joshi}}, \bibinfo {author} {\bibfnamefont
  {J.}~\bibnamefont {Krzywinski}}, \bibinfo {author} {\bibfnamefont
  {R.}~\bibnamefont {Soufli}}, \bibinfo {author} {\bibfnamefont
  {M.}~\bibnamefont {Fernandez-Perea}}, \bibinfo {author} {\bibfnamefont
  {S.}~\bibnamefont {Hau-Riege}}, \bibinfo {author} {\bibfnamefont {A.~G.}\
  \bibnamefont {Peele}}, \bibinfo {author} {\bibfnamefont {Y.}~\bibnamefont
  {Feng}}, \bibinfo {author} {\bibfnamefont {O.}~\bibnamefont {Krupin}},
  \bibinfo {author} {\bibfnamefont {S.}~\bibnamefont {Moeller}}, \ and\
  \bibinfo {author} {\bibfnamefont {W.}~\bibnamefont {Wurth}},\ }\href
  {\doibase 10.1103/PhysRevLett.107.144801} {\bibfield  {journal} {\bibinfo
  {journal} {Phys. Rev. Lett.}\ }\textbf {\bibinfo {volume} {107}},\ \bibinfo
  {pages} {144801} (\bibinfo {year} {2011}{\natexlab{a}})}\BibitemShut
  {NoStop}%
\bibitem [{\citenamefont {Singer}\ \emph {et~al.}(2012)\citenamefont {Singer},
  \citenamefont {Sorgenfrei}, \citenamefont {Mancuso}, \citenamefont
  {Gerasimova}, \citenamefont {Yefanov}, \citenamefont {Gulden}, \citenamefont
  {Gorniak}, \citenamefont {Senkbeil}, \citenamefont {Sakdinawat},
  \citenamefont {Liu}, \citenamefont {Attwood}, \citenamefont {Dziarzhytski},
  \citenamefont {Mai}, \citenamefont {Treusch}, \citenamefont {Weckert},
  \citenamefont {Salditt}, \citenamefont {Rosenhahn}, \citenamefont {Wurth},\
  and\ \citenamefont {Vartanyants}}]{Singer_FLASH_2012}%
  \BibitemOpen
  \bibfield  {author} {\bibinfo {author} {\bibfnamefont {A.}~\bibnamefont
  {Singer}}, \bibinfo {author} {\bibfnamefont {F.}~\bibnamefont {Sorgenfrei}},
  \bibinfo {author} {\bibfnamefont {A.~P.}\ \bibnamefont {Mancuso}}, \bibinfo
  {author} {\bibfnamefont {N.}~\bibnamefont {Gerasimova}}, \bibinfo {author}
  {\bibfnamefont {O.~M.}\ \bibnamefont {Yefanov}}, \bibinfo {author}
  {\bibfnamefont {J.}~\bibnamefont {Gulden}}, \bibinfo {author} {\bibfnamefont
  {T.}~\bibnamefont {Gorniak}}, \bibinfo {author} {\bibfnamefont
  {T.}~\bibnamefont {Senkbeil}}, \bibinfo {author} {\bibfnamefont
  {A.}~\bibnamefont {Sakdinawat}}, \bibinfo {author} {\bibfnamefont
  {Y.}~\bibnamefont {Liu}}, \bibinfo {author} {\bibfnamefont {D.}~\bibnamefont
  {Attwood}}, \bibinfo {author} {\bibfnamefont {S.}~\bibnamefont
  {Dziarzhytski}}, \bibinfo {author} {\bibfnamefont {D.~D.}\ \bibnamefont
  {Mai}}, \bibinfo {author} {\bibfnamefont {R.}~\bibnamefont {Treusch}},
  \bibinfo {author} {\bibfnamefont {E.}~\bibnamefont {Weckert}}, \bibinfo
  {author} {\bibfnamefont {T.}~\bibnamefont {Salditt}}, \bibinfo {author}
  {\bibfnamefont {A.}~\bibnamefont {Rosenhahn}}, \bibinfo {author}
  {\bibfnamefont {W.}~\bibnamefont {Wurth}}, \ and\ \bibinfo {author}
  {\bibfnamefont {I.~A.}\ \bibnamefont {Vartanyants}},\ }\href {\doibase
  10.1364/OE.20.017480} {\bibfield  {journal} {\bibinfo  {journal} {Opt.
  Express}\ }\textbf {\bibinfo {volume} {20}},\ \bibinfo {pages} {17480}
  (\bibinfo {year} {2012})}\BibitemShut {NoStop}%
\bibitem [{\citenamefont {Pfeiffer}\ \emph {et~al.}(2005)\citenamefont
  {Pfeiffer}, \citenamefont {Bunk}, \citenamefont {Schulze-Briese},
  \citenamefont {Diaz}, \citenamefont {Weitkamp}, \citenamefont {David},
  \citenamefont {van~der Veen}, \citenamefont {Vartanyants},\ and\
  \citenamefont {Robinson}}]{Pfeiffer_grating_2005}%
  \BibitemOpen
  \bibfield  {author} {\bibinfo {author} {\bibfnamefont {F.}~\bibnamefont
  {Pfeiffer}}, \bibinfo {author} {\bibfnamefont {O.}~\bibnamefont {Bunk}},
  \bibinfo {author} {\bibfnamefont {C.}~\bibnamefont {Schulze-Briese}},
  \bibinfo {author} {\bibfnamefont {A.}~\bibnamefont {Diaz}}, \bibinfo {author}
  {\bibfnamefont {T.}~\bibnamefont {Weitkamp}}, \bibinfo {author}
  {\bibfnamefont {C.}~\bibnamefont {David}}, \bibinfo {author} {\bibfnamefont
  {J.~F.}\ \bibnamefont {van~der Veen}}, \bibinfo {author} {\bibfnamefont
  {I.}~\bibnamefont {Vartanyants}}, \ and\ \bibinfo {author} {\bibfnamefont
  {I.~K.}\ \bibnamefont {Robinson}},\ }\href {\doibase
  10.1103/PhysRevLett.94.164801} {\bibfield  {journal} {\bibinfo  {journal}
  {Phys. Rev. Lett.}\ }\textbf {\bibinfo {volume} {94}},\ \bibinfo {pages}
  {164801} (\bibinfo {year} {2005})}\BibitemShut {NoStop}%
\bibitem [{\citenamefont {Raymer}\ \emph {et~al.}(1994)\citenamefont {Raymer},
  \citenamefont {Beck},\ and\ \citenamefont {McAlister}}]{Raymer_tomogr_1994}%
  \BibitemOpen
  \bibfield  {author} {\bibinfo {author} {\bibfnamefont {M.~G.}\ \bibnamefont
  {Raymer}}, \bibinfo {author} {\bibfnamefont {M.}~\bibnamefont {Beck}}, \ and\
  \bibinfo {author} {\bibfnamefont {D.}~\bibnamefont {McAlister}},\ }\href
  {\doibase 10.1103/PhysRevLett.72.1137} {\bibfield  {journal} {\bibinfo
  {journal} {Phys. Rev. Lett.}\ }\textbf {\bibinfo {volume} {72}},\ \bibinfo
  {pages} {1137} (\bibinfo {year} {1994})}\BibitemShut {NoStop}%
\bibitem [{\citenamefont {Tran}\ \emph {et~al.}(2007)\citenamefont {Tran},
  \citenamefont {Williams}, \citenamefont {Roberts}, \citenamefont {Flewett},
  \citenamefont {Peele}, \citenamefont {Paterson}, \citenamefont {de~Jonge},\
  and\ \citenamefont {Nugent}}]{Tran_tomogr_2007}%
  \BibitemOpen
  \bibfield  {author} {\bibinfo {author} {\bibfnamefont {C.~Q.}\ \bibnamefont
  {Tran}}, \bibinfo {author} {\bibfnamefont {G.~J.}\ \bibnamefont {Williams}},
  \bibinfo {author} {\bibfnamefont {A.}~\bibnamefont {Roberts}}, \bibinfo
  {author} {\bibfnamefont {S.}~\bibnamefont {Flewett}}, \bibinfo {author}
  {\bibfnamefont {A.~G.}\ \bibnamefont {Peele}}, \bibinfo {author}
  {\bibfnamefont {D.}~\bibnamefont {Paterson}}, \bibinfo {author}
  {\bibfnamefont {M.~D.}\ \bibnamefont {de~Jonge}}, \ and\ \bibinfo {author}
  {\bibfnamefont {K.~A.}\ \bibnamefont {Nugent}},\ }\href {\doibase
  10.1103/PhysRevLett.98.224801} {\bibfield  {journal} {\bibinfo  {journal}
  {Phys. Rev. Lett.}\ }\textbf {\bibinfo {volume} {98}},\ \bibinfo {pages}
  {224801} (\bibinfo {year} {2007})}\BibitemShut {NoStop}%
\bibitem [{\citenamefont {Alaimo}\ \emph {et~al.}(2009)\citenamefont {Alaimo},
  \citenamefont {Potenza}, \citenamefont {Manfredda}, \citenamefont {Geloni},
  \citenamefont {Sztucki}, \citenamefont {Narayanan},\ and\ \citenamefont
  {Giglio}}]{AlaimoPRL2009}%
  \BibitemOpen
  \bibfield  {author} {\bibinfo {author} {\bibfnamefont {M.~D.}\ \bibnamefont
  {Alaimo}}, \bibinfo {author} {\bibfnamefont {M.~A.~C.}\ \bibnamefont
  {Potenza}}, \bibinfo {author} {\bibfnamefont {M.}~\bibnamefont {Manfredda}},
  \bibinfo {author} {\bibfnamefont {G.}~\bibnamefont {Geloni}}, \bibinfo
  {author} {\bibfnamefont {M.}~\bibnamefont {Sztucki}}, \bibinfo {author}
  {\bibfnamefont {T.}~\bibnamefont {Narayanan}}, \ and\ \bibinfo {author}
  {\bibfnamefont {M.}~\bibnamefont {Giglio}},\ }\href {\doibase
  10.1103/PhysRevLett.103.194805} {\bibfield  {journal} {\bibinfo  {journal}
  {Phys. Rev. Lett.}\ }\textbf {\bibinfo {volume} {103}},\ \bibinfo {pages}
  {194805} (\bibinfo {year} {2009})}\BibitemShut {NoStop}%
\bibitem [{\citenamefont {Gluskin}\ \emph {et~al.}(1999)\citenamefont
  {Gluskin}, \citenamefont {Alp}, \citenamefont {McNulty}, \citenamefont
  {Sturhahn},\ and\ \citenamefont {Sutter}}]{GluskinJSynchRad1999}%
  \BibitemOpen
  \bibfield  {author} {\bibinfo {author} {\bibfnamefont {E.}~\bibnamefont
  {Gluskin}}, \bibinfo {author} {\bibfnamefont {E.~E.}\ \bibnamefont {Alp}},
  \bibinfo {author} {\bibfnamefont {I.}~\bibnamefont {McNulty}}, \bibinfo
  {author} {\bibfnamefont {W.}~\bibnamefont {Sturhahn}}, \ and\ \bibinfo
  {author} {\bibfnamefont {J.}~\bibnamefont {Sutter}},\ }\href {\doibase
  10.1107/S090904959900268X} {\bibfield  {journal} {\bibinfo  {journal} {J.
  Synchrotron Rad.}\ }\textbf {\bibinfo {volume} {6}},\ \bibinfo {pages} {1065}
  (\bibinfo {year} {1999})}\BibitemShut {NoStop}%
\bibitem [{\citenamefont {Yabashi}\ \emph {et~al.}(2001)\citenamefont
  {Yabashi}, \citenamefont {Tamasaku},\ and\ \citenamefont
  {Ishikawa}}]{YabashiPRL2001}%
  \BibitemOpen
  \bibfield  {author} {\bibinfo {author} {\bibfnamefont {M.}~\bibnamefont
  {Yabashi}}, \bibinfo {author} {\bibfnamefont {K.}~\bibnamefont {Tamasaku}}, \
  and\ \bibinfo {author} {\bibfnamefont {T.}~\bibnamefont {Ishikawa}},\ }\href
  {\doibase 10.1103/PhysRevLett.87.140801} {\bibfield  {journal} {\bibinfo
  {journal} {Phys. Rev. Lett.}\ }\textbf {\bibinfo {volume} {87}},\ \bibinfo
  {pages} {140801} (\bibinfo {year} {2001})}\BibitemShut {NoStop}%
\bibitem [{\citenamefont {Singer}\ \emph {et~al.}(2013)\citenamefont {Singer},
  \citenamefont {Lorenz}, \citenamefont {Sorgenfrei}, \citenamefont
  {Gerasimova}, \citenamefont {Gulden}, \citenamefont {Yefanov}, \citenamefont
  {Kurta}, \citenamefont {Shabalin}, \citenamefont {Dronyak}, \citenamefont
  {Treusch}, \citenamefont {Kocharyan}, \citenamefont {Weckert}, \citenamefont
  {Wurth},\ and\ \citenamefont {Vartanyants}}]{Singer_HBT_2013}%
  \BibitemOpen
  \bibfield  {author} {\bibinfo {author} {\bibfnamefont {A.}~\bibnamefont
  {Singer}}, \bibinfo {author} {\bibfnamefont {U.}~\bibnamefont {Lorenz}},
  \bibinfo {author} {\bibfnamefont {F.}~\bibnamefont {Sorgenfrei}}, \bibinfo
  {author} {\bibfnamefont {N.}~\bibnamefont {Gerasimova}}, \bibinfo {author}
  {\bibfnamefont {J.}~\bibnamefont {Gulden}}, \bibinfo {author} {\bibfnamefont
  {O.~M.}\ \bibnamefont {Yefanov}}, \bibinfo {author} {\bibfnamefont {R.~P.}\
  \bibnamefont {Kurta}}, \bibinfo {author} {\bibfnamefont {A.}~\bibnamefont
  {Shabalin}}, \bibinfo {author} {\bibfnamefont {R.}~\bibnamefont {Dronyak}},
  \bibinfo {author} {\bibfnamefont {R.}~\bibnamefont {Treusch}}, \bibinfo
  {author} {\bibfnamefont {V.}~\bibnamefont {Kocharyan}}, \bibinfo {author}
  {\bibfnamefont {E.}~\bibnamefont {Weckert}}, \bibinfo {author} {\bibfnamefont
  {W.}~\bibnamefont {Wurth}}, \ and\ \bibinfo {author} {\bibfnamefont {I.~A.}\
  \bibnamefont {Vartanyants}},\ }\href {\doibase
  10.1103/PhysRevLett.111.034802} {\bibfield  {journal} {\bibinfo  {journal}
  {Phys. Rev. Lett.}\ }\textbf {\bibinfo {volume} {111}},\ \bibinfo {pages}
  {034802} (\bibinfo {year} {2013})}\BibitemShut {NoStop}%
\bibitem [{\citenamefont {Lin}\ \emph {et~al.}(2003)\citenamefont {Lin},
  \citenamefont {Paterson}, \citenamefont {Peele}, \citenamefont {McMahon},
  \citenamefont {Chantler}, \citenamefont {Nugent}, \citenamefont {Lai},
  \citenamefont {Moldovan}, \citenamefont {Cai}, \citenamefont {Mancini},\ and\
  \citenamefont {McNulty}}]{LinPRL2003}%
  \BibitemOpen
  \bibfield  {author} {\bibinfo {author} {\bibfnamefont {J.~J.~A.}\
  \bibnamefont {Lin}}, \bibinfo {author} {\bibfnamefont {D.}~\bibnamefont
  {Paterson}}, \bibinfo {author} {\bibfnamefont {A.~G.}\ \bibnamefont {Peele}},
  \bibinfo {author} {\bibfnamefont {P.~J.}\ \bibnamefont {McMahon}}, \bibinfo
  {author} {\bibfnamefont {C.~T.}\ \bibnamefont {Chantler}}, \bibinfo {author}
  {\bibfnamefont {K.~A.}\ \bibnamefont {Nugent}}, \bibinfo {author}
  {\bibfnamefont {B.}~\bibnamefont {Lai}}, \bibinfo {author} {\bibfnamefont
  {N.}~\bibnamefont {Moldovan}}, \bibinfo {author} {\bibfnamefont
  {Z.}~\bibnamefont {Cai}}, \bibinfo {author} {\bibfnamefont {D.~C.}\
  \bibnamefont {Mancini}}, \ and\ \bibinfo {author} {\bibfnamefont
  {I.}~\bibnamefont {McNulty}},\ }\href {\doibase
  10.1103/PhysRevLett.90.074801} {\bibfield  {journal} {\bibinfo  {journal}
  {Phys. Rev. Lett.}\ }\textbf {\bibinfo {volume} {90}},\ \bibinfo {pages}
  {074801} (\bibinfo {year} {2003})}\BibitemShut {NoStop}%
\bibitem [{\citenamefont {Mej\'{i}a}\ and\ \citenamefont
  {Gonz\'{a}lez}(2007)}]{MejiaOptCommun2007}%
  \BibitemOpen
  \bibfield  {author} {\bibinfo {author} {\bibfnamefont {Y.}~\bibnamefont
  {Mej\'{i}a}}\ and\ \bibinfo {author} {\bibfnamefont {A.~I.}\ \bibnamefont
  {Gonz\'{a}lez}},\ }\href {\doibase 10.1016/j.optcom.2007.01.009} {\bibfield
  {journal} {\bibinfo  {journal} {Opt. Commun.}\ }\textbf {\bibinfo {volume}
  {273}},\ \bibinfo {pages} {428} (\bibinfo {year} {2007})}\BibitemShut
  {NoStop}%
\bibitem [{\citenamefont {Gonz\'{a}lez}\ and\ \citenamefont
  {Mej\'{i}a}(2011)}]{GonzalezJOSA2011}%
  \BibitemOpen
  \bibfield  {author} {\bibinfo {author} {\bibfnamefont {A.~I.}\ \bibnamefont
  {Gonz\'{a}lez}}\ and\ \bibinfo {author} {\bibfnamefont {Y.}~\bibnamefont
  {Mej\'{i}a}},\ }\href {\doibase 10.1364/JOSAA.28.001107} {\bibfield
  {journal} {\bibinfo  {journal} {J. Opt. Soc. Am. A}\ }\textbf {\bibinfo
  {volume} {28}},\ \bibinfo {pages} {1107} (\bibinfo {year}
  {2011})}\BibitemShut {NoStop}%
\bibitem [{\citenamefont {Goodman}(1985)}]{G1985}%
  \BibitemOpen
  \bibfield  {author} {\bibinfo {author} {\bibfnamefont {J.~W.}\ \bibnamefont
  {Goodman}},\ }\href@noop {} {\emph {\bibinfo {title} {Statistical Optics}}}\
  (\bibinfo  {publisher} {Wiley},\ \bibinfo {address} {New York},\ \bibinfo
  {year} {1985})\BibitemShut {NoStop}%
\bibitem [{\citenamefont {Saldin}\ \emph {et~al.}(2008)\citenamefont {Saldin},
  \citenamefont {Schneidmiller},\ and\ \citenamefont
  {Yurkov}}]{SaldinOptCommun2008}%
  \BibitemOpen
  \bibfield  {author} {\bibinfo {author} {\bibfnamefont {E.}~\bibnamefont
  {Saldin}}, \bibinfo {author} {\bibfnamefont {E.}~\bibnamefont
  {Schneidmiller}}, \ and\ \bibinfo {author} {\bibfnamefont {M.}~\bibnamefont
  {Yurkov}},\ }\href {\doibase 10.1016/j.optcom.2007.10.044} {\bibfield
  {journal} {\bibinfo  {journal} {Opt. Commun.}\ }\textbf {\bibinfo {volume}
  {281}},\ \bibinfo {pages} {1179 } (\bibinfo {year} {2008})}\BibitemShut
  {NoStop}%
\bibitem [{Note1()}]{Note1}%
  \BibitemOpen
  \bibinfo {note} {Here we assume that the incident intensity is constant
  across a single aperture}\BibitemShut {NoStop}%
\bibitem [{\citenamefont {Bartels}\ \emph {et~al.}(2002)\citenamefont
  {Bartels}, \citenamefont {Paul}, \citenamefont {Green}, \citenamefont
  {Kapteyn}, \citenamefont {Murnane}, \citenamefont {Backus}, \citenamefont
  {Christov}, \citenamefont {Liu}, \citenamefont {Attwood},\ and\ \citenamefont
  {Jacobsen}}]{BartelsScience2002}%
  \BibitemOpen
  \bibfield  {author} {\bibinfo {author} {\bibfnamefont {R.~A.}\ \bibnamefont
  {Bartels}}, \bibinfo {author} {\bibfnamefont {A.}~\bibnamefont {Paul}},
  \bibinfo {author} {\bibfnamefont {H.}~\bibnamefont {Green}}, \bibinfo
  {author} {\bibfnamefont {H.~C.}\ \bibnamefont {Kapteyn}}, \bibinfo {author}
  {\bibfnamefont {M.~M.}\ \bibnamefont {Murnane}}, \bibinfo {author}
  {\bibfnamefont {S.}~\bibnamefont {Backus}}, \bibinfo {author} {\bibfnamefont
  {I.~P.}\ \bibnamefont {Christov}}, \bibinfo {author} {\bibfnamefont
  {Y.}~\bibnamefont {Liu}}, \bibinfo {author} {\bibfnamefont {D.}~\bibnamefont
  {Attwood}}, \ and\ \bibinfo {author} {\bibfnamefont {C.}~\bibnamefont
  {Jacobsen}},\ }\href {\doibase 10.1126/science.1071718} {\bibfield  {journal}
  {\bibinfo  {journal} {Science}\ }\textbf {\bibinfo {volume} {297}},\ \bibinfo
  {pages} {376} (\bibinfo {year} {2002})}\BibitemShut {NoStop}%
\bibitem [{\citenamefont {{Dollas}}\ \emph {et~al.}(1998)\citenamefont
  {{Dollas}}, \citenamefont {{Rankin}},\ and\ \citenamefont
  {{McCracken}}}]{GolombProof}%
  \BibitemOpen
  \bibfield  {author} {\bibinfo {author} {\bibfnamefont {A.}~\bibnamefont
  {{Dollas}}}, \bibinfo {author} {\bibfnamefont {W.}~\bibnamefont {{Rankin}}},
  \ and\ \bibinfo {author} {\bibfnamefont {D.}~\bibnamefont {{McCracken}}},\
  }\href {\doibase 10.1109/18.651068} {\bibfield  {journal} {\bibinfo
  {journal} {IEEE Trans. Inf. Theory}\ }\textbf {\bibinfo {volume} {44}},\
  \bibinfo {pages} {379} (\bibinfo {year} {1998})}\BibitemShut {NoStop}%
\bibitem [{Note2()}]{Note2}%
  \BibitemOpen
  \bibinfo {note} {A list of all today available optimal Golomb rulers can be
  found in Ref. \cite {Golomb} (see also \cite {wiki:Golomb}).}\BibitemShut
  {Stop}%
\bibitem [{\citenamefont {{Viefhaus}}\ \emph {et~al.}(2013)\citenamefont
  {{Viefhaus}}, \citenamefont {{Scholz}}, \citenamefont {{Deinert}},
  \citenamefont {{Glaser}}, \citenamefont {{Ilchen}}, \citenamefont
  {{Seltmann}}, \citenamefont {{Walter}},\ and\ \citenamefont
  {{Siewert}}}]{JensP04_2013}%
  \BibitemOpen
  \bibfield  {author} {\bibinfo {author} {\bibfnamefont {J.}~\bibnamefont
  {{Viefhaus}}}, \bibinfo {author} {\bibfnamefont {F.}~\bibnamefont
  {{Scholz}}}, \bibinfo {author} {\bibfnamefont {S.}~\bibnamefont {{Deinert}}},
  \bibinfo {author} {\bibfnamefont {L.}~\bibnamefont {{Glaser}}}, \bibinfo
  {author} {\bibfnamefont {M.}~\bibnamefont {{Ilchen}}}, \bibinfo {author}
  {\bibfnamefont {J.}~\bibnamefont {{Seltmann}}}, \bibinfo {author}
  {\bibfnamefont {P.}~\bibnamefont {{Walter}}}, \ and\ \bibinfo {author}
  {\bibfnamefont {F.}~\bibnamefont {{Siewert}}},\ }\href {\doibase
  10.1016/j.nima.2012.10.110} {\bibfield  {journal} {\bibinfo  {journal} {Nucl.
  Instr. Meth. Phys. Res. A}\ }\textbf {\bibinfo {volume} {710}},\ \bibinfo
  {pages} {151} (\bibinfo {year} {2013})}\BibitemShut {NoStop}%
\bibitem [{\citenamefont {Vartanyants}\ \emph
  {et~al.}(2011{\natexlab{b}})\citenamefont {Vartanyants}, \citenamefont
  {Singer}, \citenamefont {Mancuso}, \citenamefont {Yefanov}, \citenamefont
  {Sakdinawat}, \citenamefont {Liu}, \citenamefont {Bang}, \citenamefont
  {Williams}, \citenamefont {Cadenazzi}, \citenamefont {Abbey}, \citenamefont
  {Sinn}, \citenamefont {Attwood}, \citenamefont {Nugent}, \citenamefont
  {Weckert}, \citenamefont {Wang}, \citenamefont {Zhu}, \citenamefont {Wu},
  \citenamefont {Graves}, \citenamefont {Scherz}, \citenamefont {Turner},
  \citenamefont {Schlotter}, \citenamefont {Messerschmidt}, \citenamefont
  {L\"uning}, \citenamefont {Acremann}, \citenamefont {Heimann}, \citenamefont
  {Mancini}, \citenamefont {Joshi}, \citenamefont {Krzywinski}, \citenamefont
  {Soufli}, \citenamefont {Fernandez-Perea}, \citenamefont {Hau-Riege},
  \citenamefont {Peele}, \citenamefont {Feng}, \citenamefont {Krupin},
  \citenamefont {Moeller},\ and\ \citenamefont {Wurth}}]{VartanyantsPRL2011}%
  \BibitemOpen
  \bibfield  {author} {\bibinfo {author} {\bibfnamefont {I.~A.}\ \bibnamefont
  {Vartanyants}}, \bibinfo {author} {\bibfnamefont {A.}~\bibnamefont {Singer}},
  \bibinfo {author} {\bibfnamefont {A.~P.}\ \bibnamefont {Mancuso}}, \bibinfo
  {author} {\bibfnamefont {O.~M.}\ \bibnamefont {Yefanov}}, \bibinfo {author}
  {\bibfnamefont {A.}~\bibnamefont {Sakdinawat}}, \bibinfo {author}
  {\bibfnamefont {Y.}~\bibnamefont {Liu}}, \bibinfo {author} {\bibfnamefont
  {E.}~\bibnamefont {Bang}}, \bibinfo {author} {\bibfnamefont {G.~J.}\
  \bibnamefont {Williams}}, \bibinfo {author} {\bibfnamefont {G.}~\bibnamefont
  {Cadenazzi}}, \bibinfo {author} {\bibfnamefont {B.}~\bibnamefont {Abbey}},
  \bibinfo {author} {\bibfnamefont {H.}~\bibnamefont {Sinn}}, \bibinfo {author}
  {\bibfnamefont {D.}~\bibnamefont {Attwood}}, \bibinfo {author} {\bibfnamefont
  {K.~A.}\ \bibnamefont {Nugent}}, \bibinfo {author} {\bibfnamefont
  {E.}~\bibnamefont {Weckert}}, \bibinfo {author} {\bibfnamefont
  {T.}~\bibnamefont {Wang}}, \bibinfo {author} {\bibfnamefont {D.}~\bibnamefont
  {Zhu}}, \bibinfo {author} {\bibfnamefont {B.}~\bibnamefont {Wu}}, \bibinfo
  {author} {\bibfnamefont {C.}~\bibnamefont {Graves}}, \bibinfo {author}
  {\bibfnamefont {A.}~\bibnamefont {Scherz}}, \bibinfo {author} {\bibfnamefont
  {J.~J.}\ \bibnamefont {Turner}}, \bibinfo {author} {\bibfnamefont {W.~F.}\
  \bibnamefont {Schlotter}}, \bibinfo {author} {\bibfnamefont {M.}~\bibnamefont
  {Messerschmidt}}, \bibinfo {author} {\bibfnamefont {J.}~\bibnamefont
  {L\"uning}}, \bibinfo {author} {\bibfnamefont {Y.}~\bibnamefont {Acremann}},
  \bibinfo {author} {\bibfnamefont {P.}~\bibnamefont {Heimann}}, \bibinfo
  {author} {\bibfnamefont {D.~C.}\ \bibnamefont {Mancini}}, \bibinfo {author}
  {\bibfnamefont {V.}~\bibnamefont {Joshi}}, \bibinfo {author} {\bibfnamefont
  {J.}~\bibnamefont {Krzywinski}}, \bibinfo {author} {\bibfnamefont
  {R.}~\bibnamefont {Soufli}}, \bibinfo {author} {\bibfnamefont
  {M.}~\bibnamefont {Fernandez-Perea}}, \bibinfo {author} {\bibfnamefont
  {S.}~\bibnamefont {Hau-Riege}}, \bibinfo {author} {\bibfnamefont {A.~G.}\
  \bibnamefont {Peele}}, \bibinfo {author} {\bibfnamefont {Y.}~\bibnamefont
  {Feng}}, \bibinfo {author} {\bibfnamefont {O.}~\bibnamefont {Krupin}},
  \bibinfo {author} {\bibfnamefont {S.}~\bibnamefont {Moeller}}, \ and\
  \bibinfo {author} {\bibfnamefont {W.}~\bibnamefont {Wurth}},\ }\href
  {\doibase 10.1103/PhysRevLett.107.144801} {\bibfield  {journal} {\bibinfo
  {journal} {Phys. Rev. Lett.}\ }\textbf {\bibinfo {volume} {107}},\ \bibinfo
  {pages} {144801} (\bibinfo {year} {2011}{\natexlab{b}})}\BibitemShut
  {NoStop}%
\bibitem [{Note3()}]{Note3}%
  \BibitemOpen
  \bibinfo {note} {At this very early stage of commissioning the grating did
  not yet deliver the designed resolving power ($E/\Delta E >
  10000$)}\BibitemShut {NoStop}%
\bibitem [{Note4()}]{Note4}%
  \BibitemOpen
  \bibinfo {note} {Our estimates show that the maximum optical path length
  difference $l$ is about $l=0.4$ $\mu $m in the region chosen for the
  analysis. This is smaller than the temporal coherence length $l_t=1.6$ $\mu
  $m and it means that our assumption $\tau < \tau _c$ is well satisfied in
  this region.}\BibitemShut {Stop}%
\bibitem [{Note5()}]{Note5}%
  \BibitemOpen
  \bibinfo {note} {In fact, for the exit slit width of 40 $\mu $m the incident
  beam was smaller than the NRA size and we additionally had to find the
  position of the incident beam relative to the center of the NRA by putting
  additional constrains such as: avoiding values of $|\gamma _{ij}|$ larger
  than one, smallest uncertainties, and smoothness of the CCF.}\BibitemShut
  {Stop}%
\bibitem [{\citenamefont {Singer}\ and\ \citenamefont
  {Vartanyants}(2011)}]{SingerSPIE2011}%
  \BibitemOpen
  \bibfield  {author} {\bibinfo {author} {\bibfnamefont {A.}~\bibnamefont
  {Singer}}\ and\ \bibinfo {author} {\bibfnamefont {I.~A.}\ \bibnamefont
  {Vartanyants}},\ }\href@noop {} {\bibfield  {journal} {\bibinfo  {journal}
  {Proceedings SPIE}\ }\textbf {\bibinfo {volume} {8141}},\ \bibinfo {pages}
  {814106} (\bibinfo {year} {2011})}\BibitemShut {NoStop}%
\bibitem [{Note6()}]{Note6}%
  \BibitemOpen
  \bibinfo {note} {This effect could be due to the fact that on the higher
  energy side of the undulator peak the collimation of the undulator cone is
  better than towards the lower energy side where at even lower energies an
  intensity minimum is observed on axis, resulting in a non-optimal beam
  path.}\BibitemShut {Stop}%
\bibitem [{\citenamefont {{Lam}}\ and\ \citenamefont
  {{Sarwate}}(1988)}]{Golomb}%
  \BibitemOpen
  \bibfield  {author} {\bibinfo {author} {\bibfnamefont {A.}~\bibnamefont
  {{Lam}}}\ and\ \bibinfo {author} {\bibfnamefont {D.}~\bibnamefont
  {{Sarwate}}},\ }\href {\doibase 10.1109/26.1464} {\bibfield  {journal}
  {\bibinfo  {journal} {IEEE Trans. Commun.}\ }\textbf {\bibinfo {volume}
  {36}},\ \bibinfo {pages} {380} (\bibinfo {year} {1988})}\BibitemShut
  {NoStop}%
\bibitem [{\citenamefont {Wikipedia}(2014)}]{wiki:Golomb}%
  \BibitemOpen
  \bibfield  {author} {\bibinfo {author} {\bibnamefont {Wikipedia}},\ }\href
  {http://en.wikipedia.org/w/index.php?title=Golomb_ruler&oldid=590065026}
  {\enquote {\bibinfo {title} {Golomb ruler --- {W}ikipedia{,} the free
  encyclopedia},}\ } (\bibinfo {year} {2014}),\ \bibinfo {note} {[Online;
  accessed 19-January-2014]}\BibitemShut {NoStop}%
\end{thebibliography}%

\end{document}